\documentclass[acmsmall]{acmart}

\AtBeginDocument{%
  }

\setcopyright{acmlicensed}
\copyrightyear{2026}
\acmYear{2026}
\acmJournal{JACM}
\acmMonth{11}
\usepackage{hyperref}
\usepackage{amsmath}
\usepackage{algorithmic}
\usepackage{graphicx}
\usepackage{caption}
\usepackage{textcomp}
\usepackage{xcolor}
\usepackage{svg}
\usepackage{afterpage}
\usepackage{booktabs}
\usepackage{float}
\usepackage{multirow}
\usepackage{todonotes}
\usepackage{soul}
\usepackage{array}
\usepackage{subcaption}
\hypersetup{
  colorlinks=true,
  linkcolor=blue,
  citecolor=blue,
  urlcolor=blue
}
\usepackage{tikz}
\usepackage{pifont}
\usepackage{rotating} 
\def\checkmark{\tikz\fill[scale=0.4](0,.35) -- (.25,0) -- (1,.7) -- (.25,.15) -- cycle;} 
\usepackage{bbding}
\usepackage[normalem]{ulem}

\begin{document}

%\title{Deep Defence on Wheels: Hybrid Quantised LSTM and Autoencoder Models based dual-IDS framework for In-Vehicle Network Security }

% \title{Deep Defence on Wheels: A Dual Intrusion Detection System Architecture for Comprehensive In-Vehicle Network Security}
\title[Deep Defence on Wheels: A Dual CAN IDS Architecture for Comprehensive In-Vehicle Network Security]
{Deep Defence on Wheels: A Dual Intrusion Detection System Architecture for Comprehensive In-Vehicle Network Security}
% \todo[inline]{Is it really a framework?}
\author{Shashwat Khandelwal}
\authornote{Both authors contributed equally to this research.}
\email{khandels@tcd.ie}
\orcid{}
\author{Shanker Shreejith}
\authornotemark[1]
\email{shankers@tcd.ie}
\affiliation{%
  \institution{Trinity College Dublin}
  \city{Dublin}
  \state{Dublin}
  \country{Ireland}
}

% \affiliation{%

\renewcommand{\shortauthors}{Khandelwal et al.}

\begin{abstract}
The combination of increasing connectivity to the outside world and the lack of inbuilt security mechanisms has made legacy intra-vehicular networks extremely vulnerable to cyberattacks.
% As legacy networks are still widely used in modern vehicles, these attacks can compromise passenger safety and data privacy by enabling attackers to manipulate critical vehicle controls and access personal information remotely.
Consequently, significant research has been devoted to analysing the impact of potential attacks, yielding datasets crucial for developing effective mitigation schemes like intrusion detection and prevention systems (IDSs/IPSs).
Initial research addressing this problem focused heavily on maximising attack detection accuracy (for both known and unknown attack vectors), often relying on large, full-precision machine learning models.
% However, embedding IDS capability within the already complex vehicular electric/electronic systems presents additional challenges beyond accuracy.
% For near real-time performance, the system is expected to support low detection latency (e.g., classification of approximately 2,000 CAN messages per second on a high-speed CAN bus).
% Additionally, the IDS should be energy efficient, minimising load on the electronic control unit’s (ECU) energy consumption, and offer minimum overhead in terms of physical resources (silicon area, memory) and processor utilisation.
However, embedding IDS into complex vehicular electronic systems adds challenges beyond accuracy, including the need for low detection latency (e.g., ~2,000 CAN frames/sec), energy efficiency, and minimal overhead in electronic control unit (ECU) resources like silicon area, memory, and processor use.
To address these deployment constraints, recent research has focused on lightweight machine learning models that balance high classification accuracy with these practical limitations.
% While some of these reduced-precision IDS models offer a good trade-off, they often fall short of achieving an optimal balance across all requirements.
In this work, we propose a dual IDS framework comprising both supervised and unsupervised learning-based solutions, each optimised for real-time, resource-constrained deployment on embedded automotive platforms. 
We present a quantised LSTM-based IDS (QLSTM-IDS) that achieves over 99.9\% detection accuracy across various attack types: DoS/Flooding, Fuzzing, and Spoofing/Malfunction, using a single model and two widely-used datasets. 
The model is trained using the brevitas quantisation aware training library, transformed into a dataflow accelerator with custom blocks compatible with AMD’s FINN toolchain, and synthesised using Vitis HLS. 
Complementing this, we introduce an 8-bit quantised convolutional autoencoder-based IDS (QCAE-IDS) for detecting previously unseen anomalies that alter CAN-ID sequence patterns, with >99\% accuracy. QCAE-IDS is quantised using AMD’s Vitis-AI toolchain. 
We detail an integration architecture enabling both IDS models to operate on a single FPGA device, bridging the network interface IP and processor system to minimise software overhead. 
QLSTM-IDS achieves 0.25ms inference latency and 0.8mJ energy consumption, while QCAE-IDS achieves 0.42ms latency and 1.1mJ energy, supporting real-time, efficient detection of both known and unknown attacks on high-speed CAN buses. 
Both solutions are deployed and evaluated on the ZCU104 SoC (XCZU7EV FPGA), demonstrating a robust, flexible hardware/software co-design for securing modern vehicular networks.
\end{abstract}




 %\received{20 November 2024}

\maketitle

\section{Introduction}
In-vehicle networks in modern vehicles enable distributed electronic control units (ECUs) to exchange control and data messages to achieve the global functions of the vehicle. 
Multiple protocols are used in vehicular systems to cater to different functions based on their criticality and to optimise the cost of electric/electronic (E/E) systems.
The Controller Area Network (CAN), due to its low cost and robustness, continues to be the standard network protocol for carrying out both critical and non-critical communication in vehicles.
Despite its popularity, CAN protocol is inherently insecure and lacks any built-in mechanisms to authenticate the transmitter, receiver, or message content~\cite{8658720}. 
CAN nodes are unable to verify the origin of messages intended for them; for instance, a data frame conveying engine stoppage information could originate from the infotainment ECU and still be accepted as valid. 
This absence of authentication renders CAN highly susceptible to attacks such as message sniffing, fuzzing, spoofing, replay attacks, and Denial-of-Service (DoS) attacks~\cite{mukherjee2016practical,enev2016automobile,koscher2010experimental,palanca2017stealth}.
Additionally, CAN utilises a broadcast-based transmission mechanism, allowing any attacker who gains control of any node on the network to send harmful frames to any other node. 
The broadcast nature also exposes CAN to passive sniffing attacks, where an attacker can surreptitiously capture all transmitted frames, including sensitive data, and subsequently use this information to execute replay attacks or more targeted spoofing attacks, which are particularly difficult to detect.
These sniffing attacks are relatively straightforward to perform, as CAN communication is unencrypted, making it easy for attackers to access and interpret critical network information.
Notably, these attacks do not require physical access to the vehicle, which can be largely attributed to the increased connectivity of vehicles to the outside world and diagnostic requirements that bridge different in-vehicle networks, which, in turn, expose internal CAN bus data to the outside world.  %\todo[inline]{is it really exposing entry points, or exposing CAN data to anyone who can access any part of the network? \textcolor{blue}{I was thinking more connectivity means more entry points which makes CAN data vulnerable? maybe like this?} - will come back to this. }
The relative ease of launching such attacks and the widespread acceptance of CAN as the protocol of choice for vehicular systems have made the detection of such attacks/intrusions an area of crucial interest in the research community for the past few years.

%The relative ease at which such attacks can be launched to compromise ECUs on critical networks 
%The occurrence of such attacks indicates potential compromise of ECUs on the critical network, making the detection of these attacks/intrusions an area of crucial interest in the research community for the past few years.

Various types of intrusion detection systems (IDSs) have been extensively explored in the research literature, including rule-based and entropy-based approaches. 
More recently, machine learning (ML) techniques based on both supervised (for detecting known attacks) and unsupervised learning (for detecting unknown attacks) have demonstrated significant success in identifying attack vectors on CAN~\cite{song2020vehicle,yang2021mth,cheng2022tcan,ma2022gru}. 
In particular, ML-based IDSs are favoured due to their generalisable nature, which has been shown to yield higher accuracy compared to traditional rule-based systems.
However, most ML-based IDSs predominantly focus on detection accuracy across various attacks as their primary evaluation metric and strive to push this metric further.
While this is an important factor, the deployment of ML-based IDSs in vehicular networks must also address other critical constraints, such as detection latency and energy consumption. 
High-speed CAN networks in critical E/E systems can exchange over 2000 messages per second on average~\cite{song2020vehicle}. 
Consequently, the IDS must be capable of processing these messages with minimal detection latency to ensure timely responses to such threats, while consuming limited energy.  
%Additionally, the IDS should be energy-efficient. 
Excessive energy consumption is another critical factor that can hinder the scalability and feasibility of IDS deployment across multiple ECUs simultaneously, thus limiting its practical application. 
Recent research has increasingly focused on these deployment challenges, leveraging techniques such as network pruning and quantisation to optimise the performance of ML models in these contexts~\cite{khandelwal2022deep, rangsikunpum2024bids, yang2021mth,de2021efficient,khandelwal2023exploring,khandelwal2025seccan}.
Given the continuously evolving vehicular applications and the deployment constraints inherent in E/E systems (such as traditional ECUs), IDS solutions should ideally be capable of performing for both known (multi-class) and unknown attack classification with high detection accuracy, preferably on a single device (such as a gateway ECU).
This feature enables the consolidation of attack detection into a limited number of ECUs, as opposed to deploying multiple IDSs for each specific attack vector and previously unseen attacks. 
However, many IDSs in the literature either lack the computational efficiency for per-message classification, cannot simultaneously detect multiple attack vectors due to limited model capacity, or are incapable of identifying unknown attacks.

In this paper, we propose a concurrent dual-IDS system for CAN comprising both a supervised and an unsupervised learning (SL \& UL) model, which has been optimised for deployment on resource-constrained vehicular ECUs. 
The SL model is a quantised Long Short-Term Memory (QLSTM) based IDS, designed for accurate, real-time per-message classification of multiple attack vectors. 
Complementing this is an 8-bit quantised convolutional autoencoder (QCAE)-based IDS, an UL solution capable of identifying unknown attacks in vehicular networks.
The QLSTM model learns ordered dependencies among the fields within individual CAN frames for supervised attack classification, while the QCAE model learns normal patterns in sequences of CAN identifiers without attack labels to identify anomalous message blocks.
The models were compressed through custom-precision quantisation and pruning for deployment on AMD’s Zynq UltraScale+ ZU7EV heterogeneous SoC, which combines FPGA programmable logic for hardware acceleration with an ARM-based processing system for system control and data movement, mimicking an extensible ECU system (for example, a gateway device). 
The precision optimisation for the models ensures a low-latency, low-energy intrusion detection architecture that can be accelerated on the FPGA fabric and coupled closely with the CAN network interface.
The following are the major contributions of the paper:
\begin{itemize}
\item We propose a quantised LSTM-based intrusion detection system (QLSTM-IDS) for CAN networks that performs per-message multiclass classification and detects multiple attack vectors using a single model.
\item We complement the supervised classification of known attacks with an unsupervised, quantised convolutional autoencoder (QCAE-IDS), trained exclusively on benign traffic to detect anomalous CAN-ID sequences that are absent from model training.
\item We present a tightly coupled hardware/software dual-IDS architecture that integrates both models with the CAN interface on a single FPGA-based SoC, thereby reducing data-movement and software overheads while satisfying the line-rate processing and energy constraints of in-vehicle deployment. To the best of our knowledge, this is among the first hardware-deployed CAN IDS frameworks to concurrently combine known-attack classification with anomalous sequence detection under real-world deployment constraints.
\item We evaluate QLSTM-IDS using two widely adopted real-world automotive attack datasets, the Car Hacking and Survival Analysis datasets on which it achieves an average classification accuracy exceeding 99.9\% across the evaluated attack vectors.
\item We evaluate the per-message detection latency and energy consumption of QLSTM-IDS against existing IDS implementations reported in the literature. Relative to the most directly comparable per-message classification approach, QLSTM-IDS achieves a 2.29$\times$ reduction in detection latency while consuming 0.8\,mJ per message-level inference.
\item We evaluate QCAE-IDS on two vehicle subsets of the Survival Analysis dataset, achieving an accuracy exceeding 99\%. Compared with our previous QCAE implementation, the proposed model maintains comparable per-block processing latency while improving energy efficiency by 1.9$\times$, with an energy consumption of 1.1\,mJ per block-level inference.
%    \item We also compare the performance of our models againtst other quantised neural networks as IDSs proposed in the research literature. 
%    \item A discussion on the trade-offs achieved by the proposed concurrent dual-IDS (QLSTM-QCAE-IDS), making them particularly well-suited for deployment in real-world vehicular environments.
\end{itemize}
%\todo[inline]{Is the QCAE tested only on Survival dataset? \textcolor{blue}{Yes, In this paper it is tested only on the survival analysis dataset.}}
Additionally, we have open-sourced the training scripts, hardware build files, and deployment scripts of the QLSTM-IDS in a publicly accessible GitHub repository \href{https://github.com/RCSL-TCD/QLSTM-IDS}{https://github.com/RCSL-TCD/QLSTM-IDS}.

The rest of the paper is organised as follows: Section~\ref{sec:background} describes the background on CAN and the various types of IDSs (SL \& UL based) proposed for it in the research literature; Section~\ref{sec:architecture} discusses the proposed models along with datasets information and their integration architecture within the ECU; Section~\ref{sec:results} discusses the results obtained and comparison with the state-of-the-art IDS/IPS proposed in the literature and Section~\ref{sec:conclusion} concludes the paper.
\section{Background \& Related works}\label{sec:background}
\subsection{Controller Area Network}
\begin{figure}[t!]
  \centering
  \includegraphics[width=\linewidth]{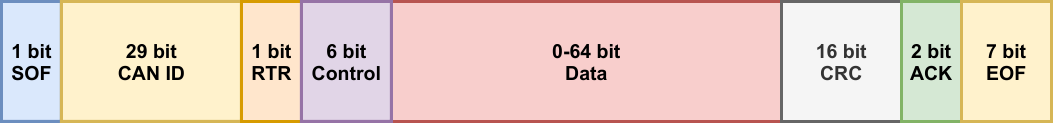}
  \caption{Frame format of an extended frame CAN message. In case of a standard CAN message, the ID field contains the 29-bit identifier.}
  \label{fig:1}
\end{figure}

CAN~\cite{CanBosch} and CAN-FD~\cite{hartwich2012can} continue to be the most widely used communication protocols in modern vehicles due to their lower cost, flexibility and robustness.
CAN bus is implemented physically using the twisted pair cable.
It is also used in other environments (e.g., controlling elevators in buildings, building energy management systems), railway applications, medical devices, and aircraft systems, among others.
Figure~\ref{fig:1} illustrates the bit-field definition of a CAN data frame.
Each segment within the frame structure provides a specific function in the network operation. 
%Each bit-field within the CAN frame serves a specific purpose: 
The start of frame (SOF) bit indicates the start of a new message, remote transmission request (RTR) bit is used to request information from another ECU, the acknowledge (ACK) field is used to indicate errors in frame and the cyclic redundancy check (CRC)  field carries a CRC value to detect bit-errors during transmission. 
Four types of CAN frames exist depending on their functionality. These are as follows:
\begin{itemize}
    \item \textit{Data frame}: The only frame among the four used for actual data transmission between nodes. The length of the payload in a CAN data frame varies from 0 to 8 bytes.
        \begin{itemize}
            \item \textit{Base frame format} : The identifier length is 11-bits.
            \item \textit{Extended frame format} : The identifier length is 29-bits (11\{base identifier\}+18\{identifier extension\}).
        \end{itemize}
    The distinction between the CAN base and extended frame formats is specified by the IDE bit, which, when transmitted as dominant (`0'), indicates a base frame and a recessive (`1') indicates an extended frame.    
    \item \textit{Remote frame}: It is also possible for a node to request information from another node in the network. This is achieved with the help of a remote frame. There are two differences between data and remote frames as follows:
        \begin{itemize}
            \item The RTR bit is transmitted as a dominant bit in a data frame and as a recessive bit in the remote frame.
            \item The remote frame does not have a payload field, which is indicated by a `0' value in the DLC field.
        \end{itemize}
    \item \textit{Error frame}: Error frames are broadcast to all nodes to indicate erroneous conditions in the network or the previous data frame. Five error types are defined in the CAN specification, namely bit error, stuffing error, CRC error, form error and ACK error. %\todo{This should be a definition of the frame type?}
    % Whenever any of these errors is detected the error frame is sent to the bus for all the nodes to drop the error message. Figure~\ref{fig:error-frame} shows the error frame format for CAN. Each node in the bus has a transmit error count (TEC) and a receive error count (REC) associated with it. The state diagram~\ref{fig:error-state-diagram} shows the node behaviour for different ranges of these values~\cite{takada2019counter}. 
    \item \textit{Overload frame}: This frame is used when a delay needs to be injected between a data frame and a remote frame.
\end{itemize}
%\todo{from here! - SS}
The CAN ID is a unique identifier assigned to each message. 
The CAN protocol uses the ID bits to define the priority of the message for transmission on the shared bus via the Carrier Sense Multiple Access with Collision Avoidance and Arbitration Priority (CSMA/CA-AP) access mechanism.
When receiving, each ECU uses the ID bits to determine if the message needs to be processed by it or ignored. %or not, since the sender(s) broadcast messages on the CAN bus. 
%It is also used as an identifier to help ECUs determine whether they should process a given message or not.
%This is necessary since the CAN traffic is broadcast in nature. 
CAN also supports multiple network speed configurations with high-speed CAN varying between 125\,kbps to 1\,Mbps (maximum speed on standard CAN) and low-speed CAN using data rates between 5\,kbps to 125\,kbps. 
CAN supports a maximum payload size of 8 bytes per message in either configuration.
This flexibility allows CAN to cater to both safety-critical functions (using high-speed CAN) and non-critical capabilities with a more cost-efficient low-speed CAN. 
CAN-Flexible Data-rate (CAN-FD) expands on standard CAN with an improved data rate to 8\,Mbps and larger payload support of up to 64 bytes, while being backwards compatible with standard CAN. 
%The maximum bitrate is 1 Mbps in the classical CAN Bus with a payload of up to 8 bytes, with the high-speed CAN Bus bitrate varying from 125 kbps to 1 Mbps and the low-speed CAN Bus bitrate from 5 kbps to 125 kbps; in the CAN Flexible Data (FD) rate, the bitrate is up to 8 Mbps with a payload size of up to 64 bytes in CAN FD. 
%While critical sensors are connected to the high-speed CAN Bus, less-critical sensors can be connected to a low-speed CAN Bus.

\subsection{IDSs for CAN}
\subsubsection{Early Works}
Researchers have explored various techniques to detect intrusions in the CAN bus.
These can be broadly classified into flow-based, payload-based, and hybrid schemes that combine the two~\cite{al2019intrusion}.
Flow-based approaches extract identifiable traits like message frequency and/or interval for the network and use these to determine abnormal activity on the network~\cite{vuong2015performance,ling2012algorithm}.
Payload-based schemes use the information in the CAN frames to detect abnormal sequences of instructions or data~\cite{stabili2017detecting,narayanan2015using}.
The hybrid schemes use both the timing information and the CAN frame fields to create a more holistic view, allowing them to extract specific signatures of transmitting ECUs, receiving ECUs, and messages~\cite{weber2018embedded,vasistha2017detecting,8688625}. 
For instance, the fingerprint-based approach uses low-level electrical signal levels and timing of signals combined with the message contents to identify potential intrusions when large deviations are observed~\cite{cho2016fingerprinting,cho2017viden}.
Parameter monitoring-based techniques extract the message-level frequency and timing characteristics for different fields or specific messages, like offset ratio and timing in case of a remote transmission frame, to identify potential intrusions~\cite{lee2017otids}.
Researchers have also proposed the use of entropy of the network traffic as a method to identify abnormal conditions and intrusions~\cite{wu2018sliding}.
While the rule-based, flow-based and hybrid techniques offer promising performance, they have been superseded by generalised machine learning techniques, including classification approaches, deep learning-based schemes, and sequential techniques, which are discussed below.

\subsubsection{Machine Learning based IDSs}
Most ML-based approaches fall under the hybrid scheme and utilise both the message content and its timing/frequency characteristics to achieve better detection. 
Tree-based approaches explored by~\cite{yang2019tree} use CAN IDs and data segment contents as input features to tree-based classifiers like decision trees, random forests, and others. 
In~\cite{song2020vehicle}, the authors propose a reduced inception net architecture for IDS, which uses deep convolutional neural networks. 
The authors show that their CNN architecture can achieve over 99\% accuracy across DoS, fuzzing, and spoofing attacks.
The authors used a dataset captured from the actual vehicle for training and testing their model, which has also been shared with the community for further research (Open Car Hacking dataset).
Since the dataset covers multiple attack modes with actual CAN messages in a real-world setting, it is one of the datasets that we used to train and evaluate our proposed model.
Recently, more complex ML architectures such as temporal convolution with global attention~\cite{cheng2022tcan}, gated recurrent unit (GRU) networks~\cite{ma2022gru}, and CNN-based temporal dependency approaches have been shown to further push the detection accuracy. 
In~\cite{10556797}, authors proposed a two-step IDS using a hybrid LSTM$+$CNN network, where the first step identifies if the message is an attack and the second step classifies the type of attack.
In~\cite{10141599}, the authors proposed a transformer network-based IDS and used batches of CAN IDs to detect attacks on multiple datasets.
While high accuracy has been obtained, the batch-based processing and high computational cost of transformer-based networks restrict their applicability for in-vehicle deployment. 
In~\cite{de2021efficient}, the authors proposed iForest, which uses an anomaly detection algorithm as an intrusion prevention system (IPS) to detect fuzzing and spoofing (RPM \& Gear) attacks and trigger the transmission of an \emph{error frame}, preventing its propagation to other ECUs.
However, with higher false positives (or in case of DoS attacks, which iForest has not been validated against), this can cause multiple messages to be dropped from the bus, leading to a partial or total loss of critical functionality.  

In addition to SL-based IDSs, multiple UL-based IDSs with the capability of detecting unknown attacks have been presented in the research literature.
In~\cite{fathima2021two}, the authors proposed the use of stacked autoencoders and GMMs as IDSs trained only on benign messages. 
However, their approach results in accuracy $<$ 80\%, making their model less suited for deployment.
In~\cite{yang2021mth}, the authors proposed a stacked anomaly-based IDS for detecting unknown attacks and reported a slightly lower average F1 score of 96.3\% on the attacks included in the CH dataset.
In~\cite{agrawal2022novelads}, the authors proposed an unsupervised learning-based CNN+LSTM autoencoder architecture for incoming message reconstruction and classifying them as benign and attacks using thresholding techniques.
The authors demonstrated that their system can achieve a high detection accuracy of 99.9\% across all attacks, while incurring higher message processing latency, making their solution unsuitable for real-time detection. 
In~\cite{anjum2022vehicle}, authors proposed the use of extreme gradient boosting machine learning algorithm to categorise unexpected occurrences in the CAN data payload.  
They reported accuracies $>$ 99\% on the SA attack dataset.
In~\cite{hossain2020lstm}, authors proposed an LSTM-based IDS to detect and mitigate the CAN bus network attacks and achieve accuracies $>$ 99.9\% on the CH and SA attack datasets.
In one of our prior works~\cite{khandelwal2023real}, we presented an UL-based convolutional autoencoder trained on the open car hacking dataset, which achieved detection accuracy of $>$ 99.5\% with near-real-time attack detection when deployed on an FPGA-based ECU.  
%Their model is deployed on an FPGA platform to enable real-time attack detection.
We arrived at a more optimal variant of the model in this work, a choice that was also inspired by other works in the literature that have explored UL-based IDS models.
%Many successful approaches presented in the literature utilise variations of autoencoders to build UL-based IDSs and hence, this model has been explored in this paper as well.

Table~\ref{table:literature-comparison} shows a comparison of these different techniques in terms of classification capabilities (known \& unknown attack detection), tested on single or multiple datasets, the input features used and whether they report deployment statistics such as processing latency and energy consumption.
In the case of the majority of works reported in the literature (for both SL- and UL-based models), inference models are accelerated on high-performance GPUs to achieve high-throughput detection, while also measuring processing latency in large batch-style operations when testing with CAN datasets. 
In contrast, our approach aims to integrate IDS accelerators as standard peripherals within a traditional ECU's dataflow and to process messages in an event-driven paradigm (when they arrive).
By compressing the models through quantisation, we can combine their high accuracy with significantly lower power consumption compared to conventional GPU-based implementations, while achieving the required throughput using clever integration methods. 

\begin{sidewaystable}[hbp]
\vspace*{12.5cm} 
\caption{The table summarises the input features utilised by various IDS/IPS approaches, along with their per-message processing capabilities, support for multi-class attack classification, evaluation across multiple datasets, ability to detect unknown attack vectors and whether they report deployment metrics such as latency and energy consumption.}
\scalebox{0.98}{
\begin{tabular}{@{}llcccccc@{}}
\toprule 
\textbf{Models} & \textbf{Input Features} & \textbf{Per-Message} &  \textbf{Multi-Class} & \textbf{Multi-Datasets} & \textbf{Unknown} &\textbf{Latency} &\textbf{Energy}     \\
\midrule
GIDS~\cite{seo2018gids}            &  ID   & \XSolid & \checkmark  & \XSolid  & \checkmark& \checkmark & \XSolid    \\
MLIDS~\cite{desta2020mlids}            &  ID   & \XSolid & \checkmark  &  \XSolid & \XSolid & \checkmark & \XSolid\\    
BIDS~\cite{rangsikunpum2024bids}            &  ID   & \XSolid  & \checkmark & \checkmark &  \XSolid & \checkmark & \checkmark \\
QdCNN~\cite{khandelwal2022deep}            &  ID   & \XSolid  & \XSolid & \XSolid &  \XSolid &\checkmark & \checkmark \\
RecCNN~\cite{desta2022rec}            &  ID   & \XSolid  & \checkmark & \checkmark & \XSolid &\checkmark & \XSolid \\
DCNN~\cite{song2020vehicle}            &  ID  & \XSolid &  \XSolid      & \XSolid &  \XSolid &\checkmark & \XSolid \\
TAN-IDS~\cite{10141599}  &  ID   & \XSolid & \checkmark & \checkmark  & \XSolid &\checkmark & \XSolid    \\
G-IDCS~\cite{park2023g}         & ID    & \XSolid & \checkmark & \checkmark & \checkmark& \XSolid & \XSolid  \\
iForest~\cite{de2021efficient}         & Payload    & \checkmark & \checkmark & \XSolid & \checkmark &\checkmark& \XSolid  \\
NovelADS~\cite{agrawal2022novelads}        &  ID + Payload  & \XSolid & \checkmark & \XSolid & \checkmark & \checkmark & \XSolid \\
XGBoost~\cite{anjum2022vehicle}        &  ID + Payload  & \checkmark & \checkmark & \checkmark &  \XSolid & \XSolid & \XSolid \\
Random Forest~\cite{santablack}        &  ID + Payload  & \XSolid & \checkmark & \XSolid  & \XSolid& \XSolid & \XSolid                     \\
CANLP~\cite{balasubramanian2024canlp}        &  ID + Payload  & \checkmark & \checkmark & \checkmark  & \XSolid& \checkmark & \XSolid \\
LSTM~\cite{hossain2020lstm}        &  ID + Payload  & \XSolid & \checkmark & \checkmark & \XSolid & \XSolid & \XSolid   \\
TCAN-IDS~\cite{cheng2022tcan}        &  ID + Payload  & \XSolid & \checkmark & \XSolid& \XSolid& \checkmark & \XSolid  \\
MTH-IDS~\cite{yang2021mth}         &  ID + Payload   & \XSolid  & \checkmark  &  \checkmark & \checkmark &\checkmark & \XSolid     \\
HybridSecNet~\cite{10556797} &  ID + Payload         & \XSolid & \checkmark &  \XSolid  &  \XSolid  &\checkmark & \XSolid \\
CQMLP-IDS~\cite{khandelwal2023exploring} &  ID + Payload         & \XSolid & \checkmark &  \XSolid & \XSolid &\checkmark & \checkmark  \\
HyDL-IDS~\cite{lo2022hybrid}        &  Time + ID + Payload + DLC  & \XSolid & \XSolid & \XSolid  & \XSolid & \XSolid & \XSolid \\
\textbf{Dual-IDS (QLSTM + QCAE)} &  ID + Payload         & \checkmark & \checkmark &  \checkmark & \checkmark &\checkmark & \checkmark   \\
\bottomrule
\end{tabular}}
\label{table:literature-comparison} 
\end{sidewaystable}

%\todo[inline]{till here}
Many research works have explored quantised neural networks (QNNs) as a way to reduce the computational complexity of ML-based IDS~\cite{khandelwal2025seccan,khandelwal2023quantised}. 
It has been shown that QNNs can maintain high attack classification accuracy (due to improved training tools) while reducing computational, memory and energy requirements significantly. 
In~\cite{rangsikunpum2024bids}, the authors introduced a hierarchical two-stage IDS based on a binarised neural network (BNN) and generative adversarial network (GAN). 
The authors demonstrated reduction in detection latency and energy consumption achieved by their IDS. 
However, their model relied solely on using CAN IDs as input features, making the IDS vulnerable to attacks that alter only the payload of CAN frames such as spoofing and replay attacks. 
In one of our previous works~\cite{khandelwal2023exploring}, we showed that quantised multi layer perceptrons (MLPs) can be deployed as a multi-class classification IDSs by utilising a conservative labelling technique to enhance model convergence across multiple attack types.

FPGA based solutions are rapidly gaining popularity in the automotive industry, due to their ability to accelerate computationally expensive algorithms for the real-time tasks through low-level optimisations, providing both energy efficiency and reconfigurability. 
Previous studies have shown their effectiveness across various applications, such as vision-based ADAS, fault detection, and high-performance Ethernet gateways~\cite{6423784,6816104,shreejith2017vega,shreejith2013approach}.
Mainstream automakers and suppliers are also exploring the viability of FPGA-based ECU platforms, with a recent demonstration implementing a real-time, heterogeneous multiprocessor system for managing motor control, battery operations, and charging functions~\cite{porschsefpgaecu}.
%Recently, Porsche has demonstrated the viability of FPGAs as part of automotive ECU platforms by integrating them into a real-time, heterogeneous multiprocessor system for managing motor control, battery operations, and charging functions. 
% FPGA complements multicore processors by handling complex real-time computations and enabling scalable, flexible ECU architectures beyond traditional microcontroller-based solutions~\cite{porschsefpgaecu}. 
% \todo[inline]{Not really sure of the last line here. Also, we are missing a link from existing works into our contribution. \textcolor{blue}{I have removed the last line here.}}

\section{System Architecture : LSTM \& CAE models }\label{sec:architecture}
\subsection{LSTM-based IDS}
% \sout{We evaluated several models proposed in the literature for use as IDSs, including Multi-Layer Perceptrons (MLPs) and Convolutional Neural Networks (CNNs). 
% However, these models either could not learn intrinsic features to perform multi-class classification or required the processing of blocks of messages with adaptations to the labelling scheme to achieve the near-line-rate performance and accuracy offered by state-of-the-art techniques in the literature. 
% Furthermore, employing more complex MLPs, CNNs or multi-stage models for multi-class classification would significantly hinder the ability to achieve near-line-rate detection, while incurring more compute resources and power consumption for the IDS function.
% To determine our specific configuration, we experimented with several \emph{n}-layer MLP architectures, we found that training smaller versions of these models to perform per-message multi-class classification was infeasible, consistent with limitations observed in prior studies~\cite{khandelwal2023exploring}.} 
% \todo[inline]{Repeating the same statement from 2 sentences ago? The key challenge is that we either need stacked models or more complex models to perform multi-class with a single message, or we need to work with message windows and label thresholding for detection? Or are we missing anything? \textcolor{blue}{Yes, this is the main challenge.}}
To arrive at a model architecture to perform multi-class classification on a per-message basis, we evaluated several architectures including MLPs and CNNs. 
We observed that large MLPs, CNNs or multi-stage configurations of these layers are required to learn the specific features of different attack vectors, which increases the computational complexity, latency and energy consumption when mapped as a dataflow model. 
On the other hand, Low-complexity models required pre-processing steps such as processing windows of messages with adapted labels and thresholding to meet similar detection performance across all attack vectors.
In general, we observed that training smaller versions of MLPs and CNNs to perform per-message multi-class classification did not converge, an observation consistent with our previous findings reported in~\cite{khandelwal2023exploring}.
We also observed that the detection performance deteriorated rapidly when applying standard model compression techniques to both low-complexity and large stacked models.

\begin{figure}[t!]
  \centering
  \includegraphics[scale = 0.4]{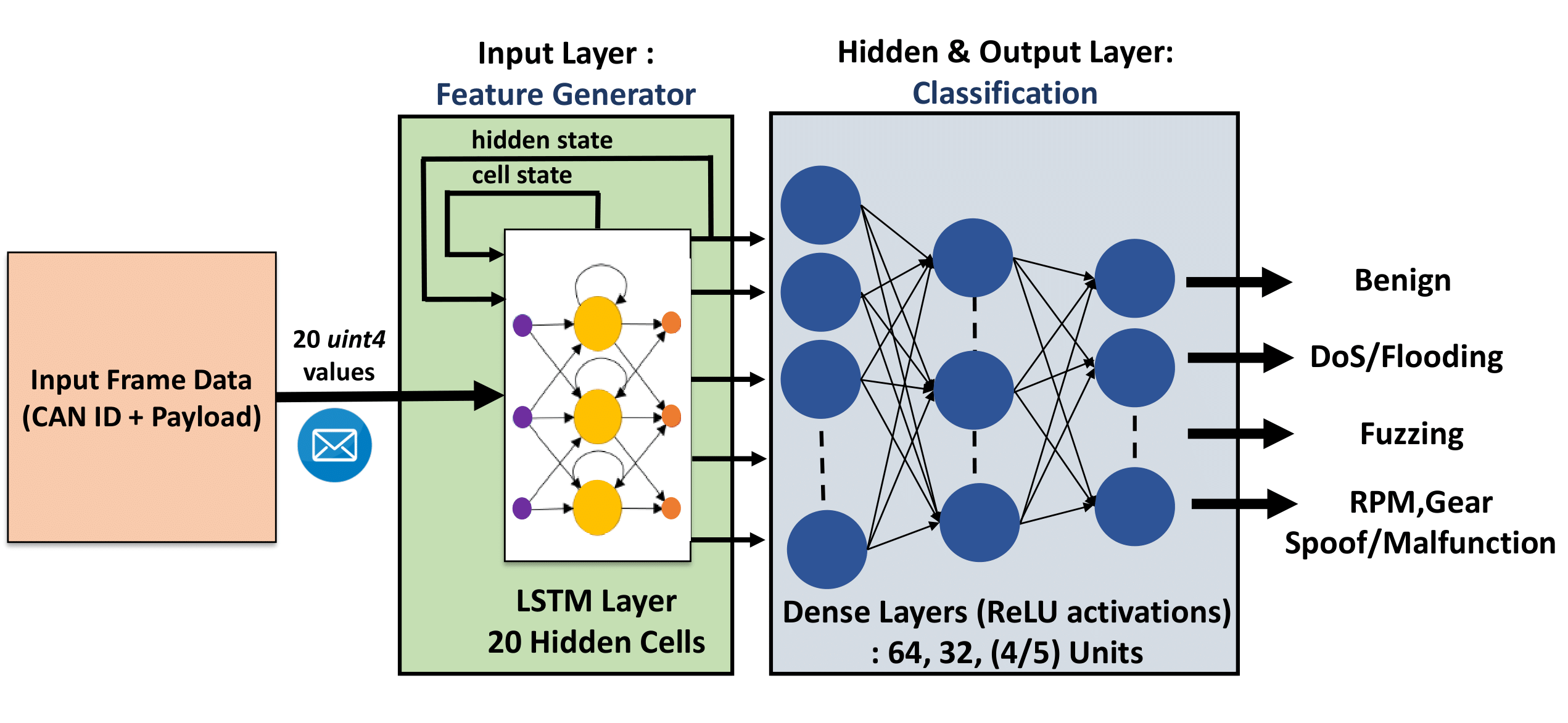}
  \caption{The figure shows the network architecture of the proposed LSTM-IDS. The CAN input is packed into 20 uint4 values as the input to the LSTM layer. The classified outputs are packed into a byte-wide data.} 
  \label{fig:network}
\end{figure}

To overcome these challenges, we explored an long short term memory network (LSTM) layer as a recurrent feature extractor over ordered segments of an individual CAN frame, followed by dense layers for multiclass classification.
Through extensive experimentation with \emph{LSTM} (feature generation) + \emph{Dense} (classification) model configurations, we observed that chaining an LSTM layer's output with a shallow dense model proved to be an ideal configuration to achieve highly accurate multi-class classification across the attack vectors. 
A high-level overview of the model's architecture is shown in Figure~\ref{fig:network}.
Each CAN frame is represented by 20 unsigned 4-bit values (\textbf{uint4}), arranged as two ordered segments of 10 features each. 
The input tensor has the shape (batch size, 2, 10), where the sequence dimension represents successive segments within a single frame rather than successive CAN messages. 
A single-layer quantised LSTM with 20 hidden units processes these two segments sequentially. 
At the first recurrent step, the layer processes the first segment; at the second step, it processes the remaining segment using the hidden and cell states produced by the first step. 
This allows the model to learn dependencies between the two portions of the current frame. 
Both states are initialised to zero for each sample in every forward pass, and no recurrent state is carried between CAN frames. 
The final hidden state therefore provides a 20-dimensional representation of the current frame alone.
The output from the LSTM layer is fed to a dense model with three fully connected layers, with 64, 32, and (4/5) neurons respectively.
The number of neurons in the last layer depend on the number of attack classes in the testing dataset.
%
%The LSTM layer is followed by a series of three fully connected (Dense) layers. 
%These layers consist of 64, 32, and 1 neurons, respectively. 
ReLU activation functions are applied between the Dense layers to introduce non-linearity, with the final output layer producing the multi-class classification result.
The final model comprises of $\approx$ 6,200 training parameters.
Accordingly, the LSTM serves as a recurrent feature extractor within individual frames, enabling per-frame multiclass classification without modelling temporal dependencies across CAN messages.
We used standard \emph{PyTorch} functions and training framework for experimenting with different configurations and for training/testing the initial models.
% Figure~\ref{fig:network} shows the model architecture of the proposed IDS.
%For initial experimentation with full-precision models, \emph{pytorch} framework was used for training/testing respectively.
%\todo[inline]{Would a network figure help here?\textcolor{blue}{Added a figure.}}

\subsection{Dataset, Training \& Deployment}
%\todo[inline]{It is not quantised yet, so QLSTM here? \textcolor{blue}{Updated}}
\subsubsection{Dataset Descriptions:} To demonstrate the effectiveness of the proposed model, we evaluate it on two distinct CAN attack datasets used widely in the research literature. 
The first dataset used is the \textbf{open Car Hacking dataset}~\cite{canlink}, which contains normal data along with separate data files for DoS, Fuzzing, and Spoofing (RPM/Gear) attacks.
Each attack file in the dataset was generated by logging data from the OBD-II port of real vehicles while targeted injections were performed at run-time. 
Each attack was sustained for 3 to 5 seconds, resulting in a dataset comprising 30 to 40 minutes of total CAN traffic. 
The attacks exhibit distinct signatures, as shown below:
\begin{itemize}
    \item \textbf{DoS Attack}: Repeatedly injecting messages with the dominant CAN ID ‘0000’ every 0.3 milliseconds to overwhelm the network.
    \item \textbf{Fuzzing Attack}: Injecting messages with completely random CAN ID and DATA values every 0.5 milliseconds to disrupt normal network operation.
    \item \textbf{Spoofing Attack (RPM/Gear)}: Injecting messages with specific CAN IDs associated with RPM or gear information every 1 millisecond to imitate or alter legitimate data.
\end{itemize}
% \sout{To truly test the multi-class classification capabilities of our model, we combined all attack data files for the training, validation, and testing phases.}\todo{we specify this later, so might not need it here?\textcolor{blue}{Okay!}} 

The second dataset we used to evaluate our model and demonstrate its generalization capability is the \textbf{Survival Analysis Dataset for Automotive IDS}~\cite{han2018anomaly}.
This dataset primarily focuses on three types of attacks that can severely disrupt in-vehicle functions or escalate attack impact and damage: Flooding, Fuzzing, and Malfunction attacks.
Like the Car Hacking dataset, this dataset includes one file for normal traffic and three separate files for each attack. 
The attack packets are injected for 5 seconds at 20-second intervals for each type of attack.
To enhance generalizability, these three attacks were carried out on three different vehicles: the Hyundai Sonata, Kia Soul, and Chevrolet Spark. 
The specific attack signatures within this dataset are as follows:
\begin{itemize}
    \item \textbf{Flooding Attack:} This attack resembles the DoS attack in the Car Hacking dataset, with both datasets injecting dominant CAN ID 0x0000. 
    However, this dataset introduces complexity by including multiple message IDs with payloads set to all zeros across all 8 bytes, making the attack pattern slightly more intricate.
    \item \textbf{Fuzzing attack:}  Similar to the Car Hacking dataset, this attack involves injecting messages with random CAN IDs and payloads on the bus every 0.3 milliseconds.
    \item \textbf{Malfunction attack:}  The malfunction attack targets specific CAN IDs similar to the spoofing attack in the Car Hacking dataset. In this attack, manipulation of the data field is combined with the injection of messages using randomly selected CAN IDs to disrupt the vehicle's functionality. IDs specific to each vehicle were chosen to launch the attack. %CAN IDs 0×316, 0×153, and 0×18E were selected from the Hyundai Sonata, Kia Soul, and Chevrolet Spark respectively for this attack. 
\end{itemize}
Both these datasets have been widely used in the research literature for developing and evaluating various IDS approaches. 
Consequently, we leverage these datasets to rigorously test, validate and demonstrate the effectiveness of our IDS model.

We pre-process both datasets to filter out features (CAN Packet Data) we used for our training and testing flows.
All the attack datasets were also preprocessed to convert the CAN ID $+$ Payload features used as inputs to our model to the \textbf{uint4} datatype as this datatype allowed us to retain the highest accuracy through the model compression stages. % us the best model accuracy as discussed in the latter sections.
For the Car Hacking Dataset, the training set consisted of 10 million CAN frames (84.3\%), the validation set contained 1.24 million CAN frames (10.4\%), and the test set included 620,200 CAN frames (5.2\%). 
These subsets were constructed by selecting an equal number of messages from the dataset's four attack categories: DoS, Fuzzing, RPM-Spoof, and Gear-Spoof.
In the case of the Survival Analysis Dataset, all datasets captured from the same car model were combined to train the model for multi-class classification. 
However, we observed that the number of CAN frames captured from the three vehicles (as provided by the dataset) varied significantly. 
To ensure balanced datasets between benign and attack messages, the following train, validation, and test splits were applied for each model:
\begin{itemize}
    \item Sonata model: 260,000 CAN frames (67.5\%) for training, 50,000 CAN frames (12.9\%) for validation, and 75,000 CAN frames (19.4\%) for testing.
    \item Soul model: 360,000 CAN frames (70.5\%) for training, 60,000 CAN frames (11.7\%) for validation, and 90,000 CAN frames (17.6\%) for testing.
    \item Spark model: 172,000 CAN frames (67.7\%) for training, 32,000 CAN frames (12.5\%) for validation, and 50,000 CAN frames (19.6\%) for testing.
\end{itemize}
Due to the relatively smaller size of the Survival Analysis Dataset, a larger proportion of the data was allocated to testing to effectively demonstrate the efficacy of the proposed model’s training and performance.

\subsubsection{Training procedure}
\begin{figure}[t]
    \centering
    \begin{subfigure}[b]{0.49\textwidth}
        \includegraphics[width=\textwidth]{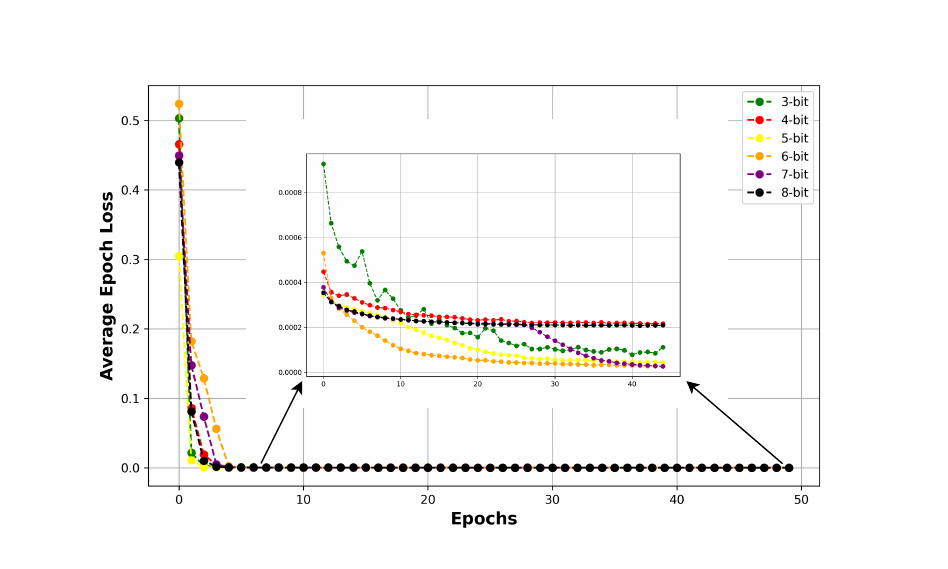}
        \caption{Average Training Loss : DSE}
        \label{fig:subfig1}
    \end{subfigure}
    \hfill
    \begin{subfigure}[b]{0.49\textwidth}
        \includegraphics[width=\textwidth]{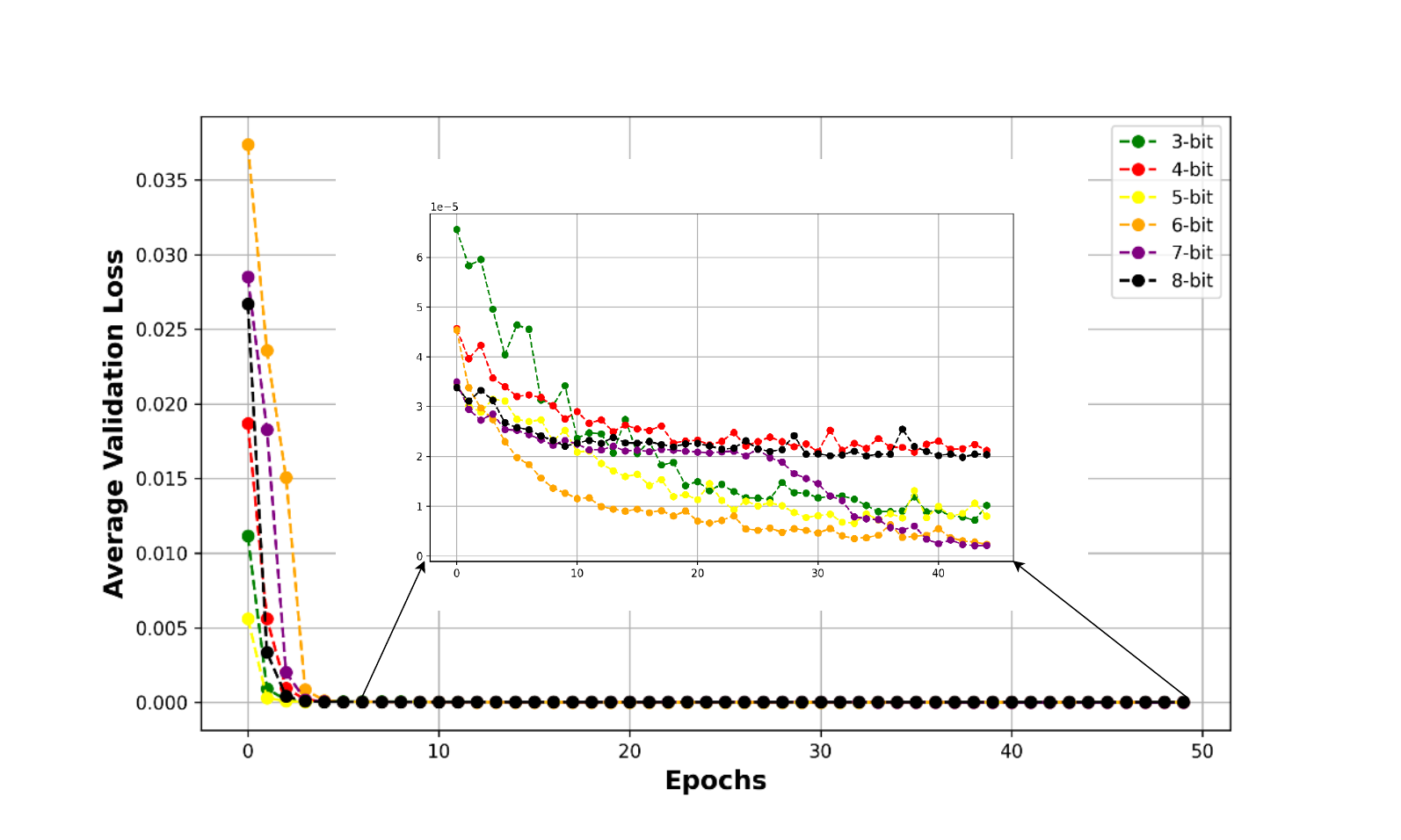}
        \caption{Average Validation Loss : DSE}
        \label{fig:subfig2}
    \end{subfigure}
    \caption{The figure illustrates the average training and validation loss curves for the QLSTM-IDS during its training for different activation precisions using the car hacking dataset. To better highlight the differences, the figure provides a magnified view of epochs 5 to 50.}
    \label{fig:training-plots-dse}
\end{figure}

\begin{figure}[t]
    \centering
    \begin{subfigure}[b]{0.49\textwidth}
        \includegraphics[width=\textwidth]{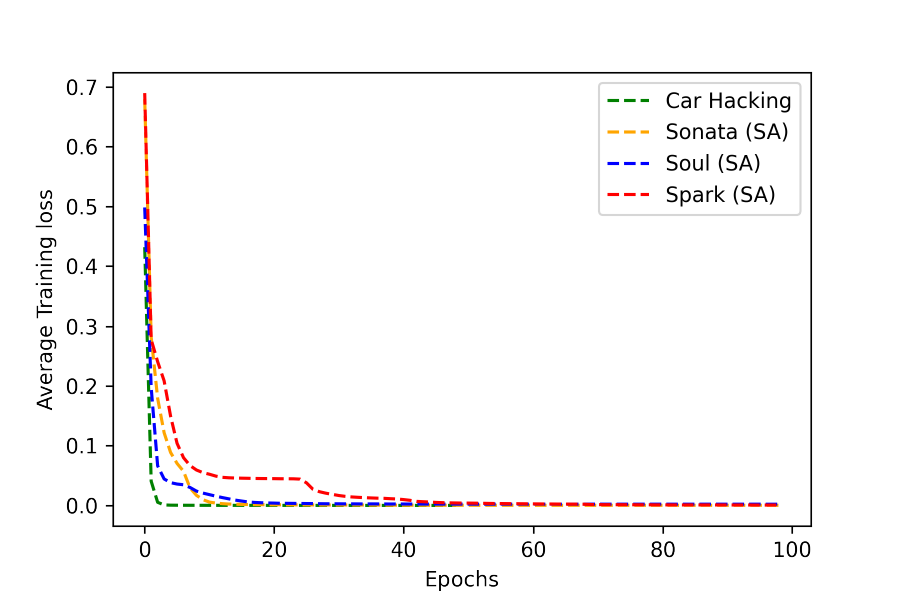}
        \caption{Average Training Loss}
        \label{fig:subfig1}
    \end{subfigure}
    \hfill
    \begin{subfigure}[b]{0.49\textwidth}
        \includegraphics[width=\textwidth]{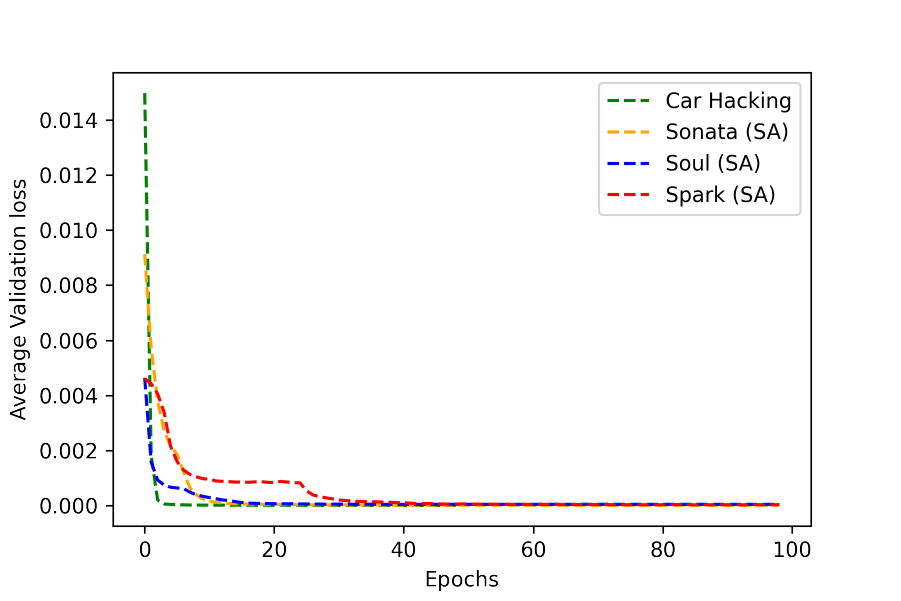}
        \caption{Average Validation Loss}
        \label{fig:subfig2}
    \end{subfigure}
    \caption{The figure illustrates the average training and validation loss curves for the QLSTM-IDS during its training on the four attack datasets from the car hacking and survival analysis (SA) datasets.}
    \label{fig:training-plots}
\end{figure}

After arriving at the model architecture (LSTM $+$ Dense) by observing its efficacy at floating point precision (fp32), we explored mixed-precision quantisation to reduce the model's compute and memory requirements. 
We quantised the LSTM and Dense layers from fp32 weights/activations to a quantised version with 8-bit quantised weights and 6-bit quantised activations (including the internal activations of the LSTM layer).  %\todo[inline]{Only LSTM layer? what about Dense?}
A design space exploration (DSE) on the model's weight bit-widths was not conducted, as the corresponding matrix multiplication operations into which these weights are synthesised in hardware do not significantly impact resource utilisation. 
However, a DSE for activation bit-widths was essential, as resource usage increases exponentially with activation precision. 
This is because activations are implemented as Multithreshold blocks in hardware, which scale non-linearly with the activation bit-width~\cite{umuroglu2017streamlined}.
We train the quantised version of this model with \emph{brevitas}, an open-source quantisation-aware training (QAT) library by AMD. 
% \todo[inline]{did you not have to do specific optimisations to brevitas to support LSTM? I met Nicolas the other day, and he was mentioning the changes you had to make, which is now part of the upstream Brevitas code, so we could refer to those here? \textcolor{blue}{I don't remember optimizations, The only thing that I did was to add support for the export of brevitas QLSTM model as an ONNX graph. I have added lines in blue talking about this in section 3.2.3}}
For all the datasets, we used the \emph{Adam} optimiser with the \emph{CrossEntropyLoss} as the loss function for multi-class classification. 
The \emph{CrossEntropyLoss} loss function was chosen as they are known to improve the training flow when dealing with a multi-class classification problem with unbalanced training set like in our case.
% \sout{(known to be useful when training a classification problem with \emph{k} classes and particularly useful when dealing with unbalanced training set like we have) for multiclass attack classification.}
We set the \emph{learning rate} for all our training flows to \emph{0.0001}.
A slower \emph{lr} helps to improve model convergence as has been shown by many studies~\cite{wu2018training}.
With the car hacking dataset, the model converged quicker at \emph{50} epochs with a batch size of \emph{2000}.
In the case of the survival analysis dataset, the convergence was slower due to the smaller training set, requiring the model to be trained for \emph{100} epochs with a batch size of \emph{200} to achieve the best accuracies. 
Dropout layers were incorporated between computational layers to mitigate overfitting, enhancing the model's generalisation and training efficiency. 
Additionally, weight clipping techniques were employed during the training to the address gradient explosion in the LSTM layer, ensuring stable gradient updates and improved convergence.
Figure~\ref{fig:training-plots-dse} presents the average training and validation loss for activation bit-widths ranging from 2 to 8. 
The 2-bit activation model was excluded from the plot, as it exhibited overfitting during training. 
Ultimately, 6-bit activations were selected with a combination of 8-bit weights, offering the lowest validation loss across all configurations trained for 50 epochs.
Figure~\ref{fig:training-plots} presents the training and validation loss curves for the QLSTM-IDS model trained on the four attack datasets. 
The training process exhibits stable convergence, with the best-performing models for each dataset (Car Hacking, Sonata (SA), Soul (SA), and Spark (SA)) selected at epochs \emph{29}, \emph{64}, \emph{75}, and \emph{83}, respectively, based on the lowest validation loss.

\subsubsection{Generating the dataflow accelerator}
Generating the dataflow hardware model for generalised LSTM models are not natively supported by open-source tools such as \emph{FINN}. 
However, it has been shown that custom implementations of uniformly quantised LSTMs can be achieved using hardware building blocks available in FINN's high-level synthesis library (\textit{finn-hlslib}) as shown in~\cite{rybalkin2018finn}.
%Generalised LSTM deployment on FPGAs are currently not supported with \emph{FINN} (toolchain to build hardware for QNNs).
%However, custom implementation of LSTMs using hardware building blocks available in \href{https://github.com/Xilinx/finn-hlslib}{\textit{finn-hlslib}} have been proposed in the research literature~\cite{rybalkin2018finn}.
We take inspiration from this approach and convert our trained mixed precision QLSTM-Dense model to the \emph{ONNX} intermediate representation (IR).
Brevitas export flow does not have support for quantised LSTM layers, for which we built a custom handler function to generate the ONNX model for our QLSTM model.
%\textcolor{blue}{This export to ONNX format especially for the QLSTM layer is not part of brevitas export flow yet and we had to update certain handler functions in this flow to make this export possible.}
This export representation captures the computations, weights as well as scales/zero-points of the quantisers that we used to achieve the quantisation, which is required by the mapping flow to generate the hardware dataflow block. 
%
%This representation acts as the starting point for us \textcolor{blue}{(exporting weights as well as the scales/zero-points of the quantizers used to achieve quantisation)} which we used to generate the dataflow hardware accelerator for the QLSTM-Dense model.
To map the ONNX nodes to the available hardware building blocks, we reverse-engineered certain transformations provided by FINN and created new transformations in the FINN compiler to map the computations in our ONNX graph to finn-hlslib hardware blocks.
In this step, we transformed graph nodes that do not have a one-to-one representation in the library using the custom transformations to represent them using FINN-compatible hardware blocks. 
Subsequently, the updated computation graph was passed to Vitis HLS (v2022.2) flow and linked with the FINN libraries to map these blocks using the hardware blocks available in the FINN backend. 
Pragma's were used to unfold loops for parallel processing as well as to map the input and output interfaces of the model as AXI stream allowing for seamless integration with other IP blocks and/or processing system on a Zynq device (the hardware description script can be accessed via our \href{https://github.com/RCSL-TCD/QLSTM-IDS/tree/main/backend}{\underline{GitHub repository}}).
%
%The following the steps of the previous custom LSTM implementation with FINN tools, we emulated the transformed graph using FINN-compatible hardware blocks and described the final computation graph in Vitis HLS (v2022.2) using hardware blocks available in the FINN backend to generate the final IP block for the hardware accelerator.
For our testing, the IP was integrated into a PYNQ overlay design with a data-preprocessor FIFO which converts byte-wide CAN message data into the \textbf{uint4} representation used by the model.
The PYNQ overlay includes AXI-DMA and other interface wrappers to move data from/to the IP from software on the ARM cores, to aid with our evaluation. 
%This IP block is integrated into a larger design as a PYNQ overlay, including AXI-DMA and other components to enable data read/write operations from the IP.
The Zynq Ultrascale+ XCZU7EV device on the ZCU104 development platform was chosen as a target platform for the implementation. 
For testing the I/O interface and the final model, the PYNQ interface was used to write data to the model via the pre-processor FIFOs and to read the classified result from the accelerator IP.
This flow was chosen as it allows us to determine the classification accuracy, detection latency and energy consumption of the hardware implementation of our model while embedding the IDS function in an architecture that mimics a standard ECU. 
While the PYNQ flow introduces software overheads, it provides a seamless interface for moving data between the ARM cores and logic and provides API functions to probe on-board power rails directly from software at runtime~\cite{amd_pynq_2025}.
%\todo[inline]{Have to think what to add here as it should not contradict with the next integration scheme}
%Finally, we test the IP by using PYNQ read/write operations and note the classification accuracy, detection latency and live energy consumption of the QLSTM-IDS model on the ZCU104 development board.

\subsection{CAE based IDS}~\label{subsec:cae-ids}
%\textcolor{blue}{%
While the LSTM-based IDS discussed above is trained to detect known attack vectors, it remains vulnerable to novel or evolving threats that fall outside its training distribution, a common limitation of supervised models. 
In prior work, we proposed an autoencoder-based IDS that learns the normal operational patterns of CAN bus traffic and flags deviations as potential anomalies~\cite{khandelwal2023real}. 
This model was trained exclusively on benign data from the CH dataset and evaluated on sequences that combined benign and malicious traffic.
In this work, we streamline and optimise that model architecture, and evaluate its performance on a distinct dataset: the SA attack dataset, specifically the Sonata and Spark subsets.
The model architecture, input feature size parameters and training/post-quantisation fine-tuning steps were optimised to learn the normal message patterns from the smaller training data corpus (9$\times$ fewer samples compared to our previous work). 
Note, in this work, “unseen attacks” denotes attacks excluded from training, model selection, and threshold calibration and introduced only during final evaluation. 
These attacks belong to established categories in the SA dataset; they are unseen by the model rather than novel attack categories or demonstrations of real-world zero-day exploitation. 
The evaluation therefore assesses benign-trained anomaly detection for attacks that alter CAN-ID sequence patterns.
% in these datasets for learning normal patterns, 
%Despite the limited training data available (fewer than 9× fewer samples), we were able to adapt and optimise our previous implementation to perform \textbf{unseen attack detection} effectively on this dataset. 
%Our results show that the autoencoder-based model maintains high detection accuracy despite being trained on different data, demonstrating its capacity to generalise across diverse vehicle platforms and attack profiles.
%} \todo[inline]{we cannot be making conclusions here - this section is more about the architecture?\textcolor{blue}{There are no conclusions I think in this paragraph. The architecture, optimal block and threshold determination is discussed here.}}
%\todo[inline]{should we say we streamlined the model further and explore its generalisability? Otherwise, it feels like it is a reuse of existing work, so not novel? Also, is this only on the SA dataset? And if so, is it still considered generalised? \textcolor{blue}{I updated the paragraph based on this comment.}}
 
\begin{figure*}[t!]
    \centering
    \includegraphics[scale = 0.65]{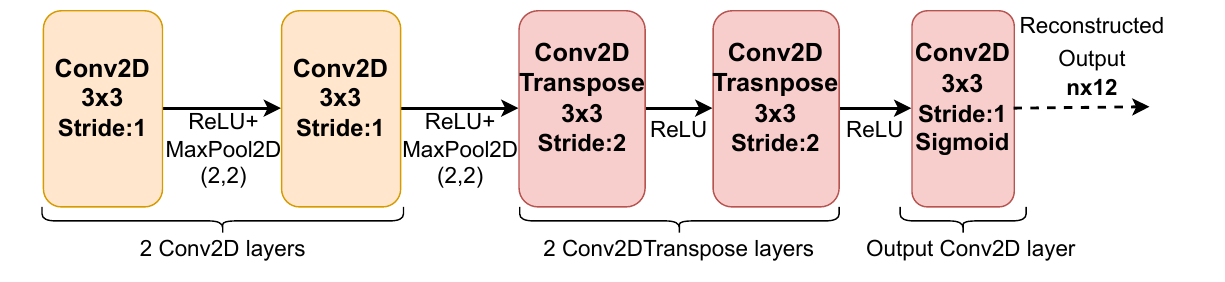}
    \caption{The proposed convolutional autoencoder model architecture for CAN-ID sequence anomaly detection IDS. `n' denotes the window size of the input feature used for attack detection. The proposed model uses $n=100$ in this paper.}
    \label{fig:zerodaymodel}
\end{figure*}

The architecture of the autoencoder is based on an encoder-decoder framework. 
The encoder consists of two \texttt{Conv2D} layers with \texttt{ReLU} activation, and max-pooling layers are inserted between the convolutions to reduce the feature dimensions progressively. 
The decoder is composed of two \texttt{Conv2DTranspose} (Deconvolution) layers, also using \texttt{ReLU} activations, to reconstruct the input from the learned latent space. 
The final layer of the decoder consists of a \texttt{Conv2D} layer with a single filter and \texttt{sigmoid} activation, producing the binarised reconstructed message. 
This model architecture is shown in Figure~\ref{fig:zerodaymodel}.
The model was implemented in Tensorflow, using standard TensorFlow functions and node definitions.
As the input feature, a sequence of \texttt{n = 100} CAN-IDs is chosen in binary format while experimenting with the dataset, capturing the sequence of messages exchanged on a normal CAN network. 
At a bus utilisation of 100\%, this corresponds to a time window of 1 ms (with the peak rate of 10,000 messages per second) on a high-speed CAN network.
During the DSE, it was observed that using smaller blocks (e.g., \(n = 10\) or \(n = 50\)) limited the model's ability to learn intrinsic features, reducing the classifier's efficiency. 
On the other hand, increasing the window size (i.e., block size \(n > 100\)) resulted in much higher worst-case detection latency, as more time was required to accumulate the required number of messages for attack detection and hence  \(n = 100\) was chosen as the block size for this dataset.

\begin{figure}[t!]
  \centering
  \includegraphics[scale = 0.575]{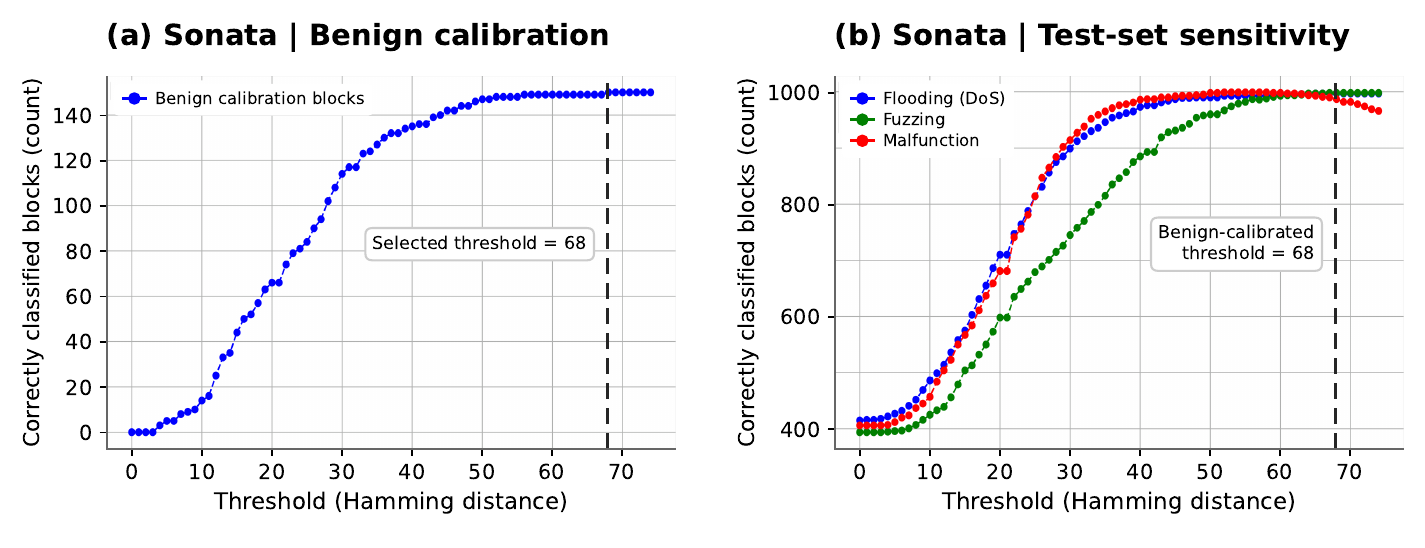}
  \caption{Floating-point CAE threshold calibration and test-set sensitivity for the Sonata subset. (a) Correctly classified benign calibration blocks out of 150, with a selected threshold of 68. (b) Correctly classified blocks across thresholds for the Flooding, Fuzzing, and Malfunction test sequences. The dashed line marks the threshold determined exclusively from benign calibration data; test-set results did not inform its selection. Zero calibration false positives do not constitute an independent test result.}
  \label{fig:side_by_side_attack_comparison}
\end{figure}

\begin{figure}[t!]
  \centering
  \includegraphics[scale = 0.575]{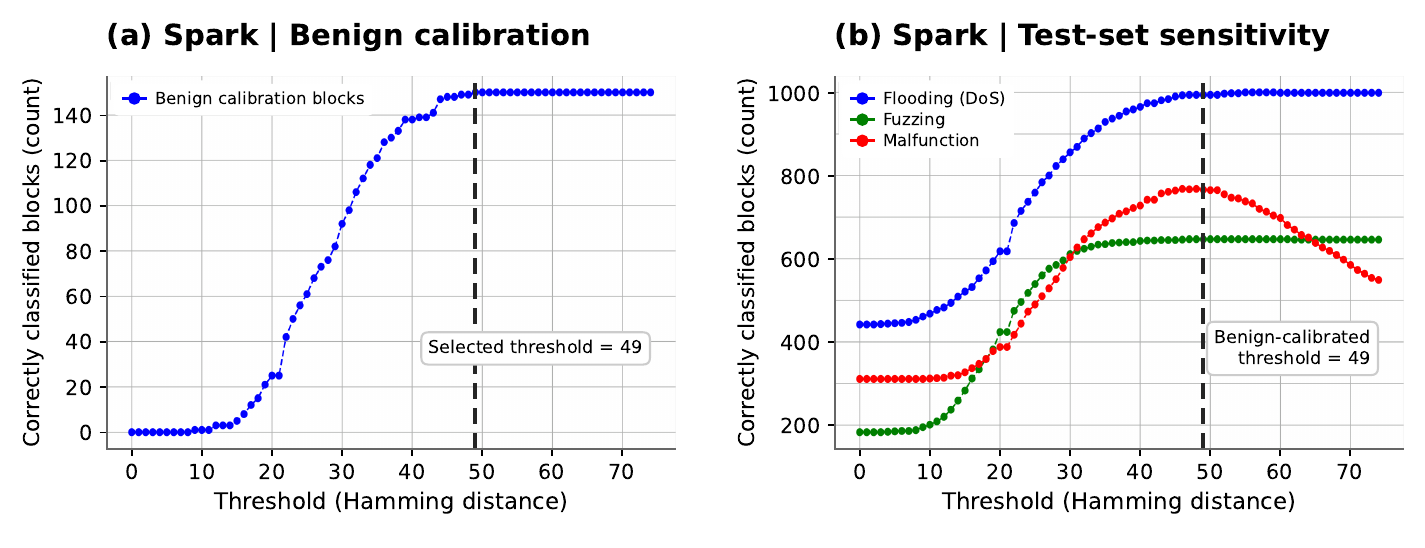}
  \caption{Floating-point CAE threshold calibration and test-set sensitivity for the Spark subset. (a) Correctly classified benign calibration blocks out of 150, with a selected threshold of 49. (b) Correctly classified blocks across thresholds for the Flooding, Fuzzing, and Malfunction test sequences, containing 1,000, 650, and 790 blocks, respectively. The dashed line marks the threshold determined exclusively from benign calibration data; test-set results did not inform its selection. Zero calibration false positives do not constitute an independent test result.}
  \label{fig:side_by_side_attack_comparison_spark}
\end{figure}

To determine the optimal configuration of the CAE, different design configurations with different layers and filter sizing were explored, iterating through multiple training and validation steps.
For the \texttt{Conv2D} and \texttt{Conv2DTranspose} layers different filter sizes \{128,\,64,\,32\} both in the encoder and decoder were explored. 
For the SA dataset, we found that the configuration \{64,\,32\} \& \{32,\,64\} for the filter sizes in the encoder and decoder gave the highest classification performance (as opposed to the \{128,\,64\} \& \{64,\,128\} configuration achieved for the CH dataset~\cite{khandelwal2023real}). 
% Due to the significantly lower volume of normal message data (less than 9 $\times$) compared to the CH dataset, the smaller model converges more quickly and prevents overfitting when employing the larger model architecture.
To classify the reconstructed messages as benign or an attack, the \emph{hamming distance} metric was chosen to compute the distance (threshold) between the original and predicted binary vectors.
With the network trained on normal ECU communication, any abnormal or attack message flow in the input stream causes the reconstructed message to have a much larger hamming distance (greater than the classification threshold) as the model has not seen these messages during the training phase.
To determine the classification threshold, the benign dataset for each vehicle was divided into separate training, validation, and threshold-calibration subsets. 
The QCAE model weights were optimised using the training subset, and the final model checkpoint was selected according to the lowest loss on the validation subset. 
After fixing the model weights, the reconstruction-error threshold was varied from 0 to 100 using only the disjoint benign calibration subset. 
For the Sonata model, the number of false positives decreased as the threshold increased from 0 to 67. 
A threshold of 68 was the smallest value that produced no false positives on the benign calibration subset and was therefore selected. 
The same procedure produced a threshold of 49 for the Spark model. 
No attack samples were used for model training, checkpoint selection, or threshold calibration. 
After the model and threshold had been fixed, the separate attack datasets were used exclusively for evaluating the detection of attack traffic absent from training.
Figures~\ref{fig:side_by_side_attack_comparison} \& ~\ref{fig:side_by_side_attack_comparison_spark} present the threshold-calibration results and the subsequent attack-evaluation results for the Sonata and Spark subsets, respectively.

The anomaly-detection capability of QCAE is limited by its input features. 
As the model processes only sequences of CAN IDs, it is primarily sensitive to anomalies that change the occurrence, ordering, or frequency of identifiers, including flooding, fuzzing, and message-injection attacks that disturb normal CAN-ID patterns. 
It may not detect attacks that modify only the payload while preserving an otherwise normal identifier sequence, such as certain payload-spoofing, replay, or semantic-manipulation attacks. 
The QLSTM component provides complementary payload-aware detection because it processes both CAN IDs and payload data. 
It can therefore identify payload manipulations that resemble attack classes represented during its supervised training. 
Nevertheless, the current architecture does not guarantee detection of previously unseen payload-only attacks, and extending QCAE to incorporate payload or timing features remains an important direction for future work.

\subsubsection{Training Procedure} As mentioned, Sonata \& Spark subsets of the SA dataset were used for training the model and to test the performance~\cite{han2018anomaly}. 
The attack free datasets in each of the subsets were split in a 75:15:10 ratio for training, validation, and threshold-calibration steps respectively, with the largest section allocated to training and optimisation of the quantised network. 
For the Sonata subset, $100,000$ messages were extracted in sequence from each of the attack sets, resulting in $1000$ blocks \(n = 100\) for testing.
For the Spark subset, $100,000$, $65,000$ \& $79,000$ messages were extracted in sequence from each of the attack sets, resulting in $1000$, $650$ \& $790$ blocks (\(n = 100\)) for testing for each attack, respectively. %\todo[inline]{What does these lines mean? \textcolor{blue}{They just tell how the normal message sets for both subsets were created. And then how were the attack sets pre-processed and how many blocks they each comprise of. It is slightly messy as each set had different amount of messages and I tried to take the maximum out of them.}}
The models performance on the validation set during training ensures that it is not overfitting on the benign message dataset. 
% The subsets were pre-processed (as blocks of 100 messages) before training to mimic the CAN data format that the model will obtain as its input when integrated into the ECU. 
With 12-bit CAN IDs used in the dataset, the binarised input block has the shape \texttt{\{100,12\}}, which is organised as the input FIFO/buffer when integrated with the PS.  
These blocks are then fed into the model as input for training and testing, respectively.
The models were trained using the standard TensorFlow framework.
The \texttt{Adam} optimiser was used with \texttt{binary cross entropy} as the loss function during training.
% The BCELoss function is given by:
% \begin{equation}
%     Loss =
% L = - \frac{1}{N} \sum_{i=1}^{N} \left[ y_i \log(p_i) + (1 - y_i) \log(1 - p_i) \right]
% \end{equation}
The learning rate for training the models was set to $0.001$ and they comprised of 47,425 learnable parameters. %\todo[inline]{Doesn't make sense here. Is this the case for the final model? Because in the next line, you mention different models were trained, and the best model is chosen?\textcolor{blue}{So there are two models here for two different datasets but both of the them have the same architecture and were trained in a similar way.}}
The models were trained for 500 epochs with a batch size of 64 on the normal (attack-free) dataset.
The best model in terms of validation loss was saved and exported as an \texttt{'h5'} file.
The weights file for the models were then fed into the Vitis-AI toolchain for quantisation and post-quantisation optimisations for final deployment on the DPU IP.

\subsubsection{Generating the overlay accelerator}
Once the proposed CAE's architecture and parameters were determined through DSE, the weights, biases, and activations of the trained model are 8-bit quantised using AMD/Xilinx's Vitis-AI framework to generate the quantised versions of the CAE model (referred to from hereon as QCAE-SA)~\cite{xilinxvitis}. %\todo[inline]{You are jumping between past and present tense too much. Previously it `was' trained, `were' exported etc, and now you `are' quantising...}
Following quantisation of the floating-point models using Vitis-AI, the reconstruction-error thresholds were recalibrated using the same benign-only procedure applied to the floating-point models. 
Quantisation changes the numerical representation of the model parameters and activations and can therefore alter the reconstruction errors, requiring the thresholds to be determined again for the quantised models. 
With the quantised models fixed, the threshold was varied using only the benign threshold-calibration subset, and the smallest value yielding no false positives on that subset was selected. 
This procedure resulted in thresholds of $73$ and $52$ for the Sonata and Spark models, respectively, compared with the floating-point thresholds of $68$ and $49$. 
No attack samples or test-set results were used to select or adjust the thresholds at either stage. 
The final post-quantisation thresholds were then frozen and used for subsequent evaluation on the respective test datasets.
%Note, the threshold post-quantisation for the QCAE-SA model on the Sonata and Spark models, shift from $67$ to $73$ and $49$ to $52$, which are then used to verify the functional correctness of that model on their respective test sets.
To achieve optimal acceleration of the model, different hardware accelerator configurations of the Vitis-AI's DPU IP core were explored (B512, B1152, B4096), taking into account the resource constraints on the target Ultrascale+ (XCZU7EV) platform~\cite{xilinxdpu}. 
These DPU configurations were evaluated to compare the latency and energy consumption when implementing the models, and we observed that the B512 DPU provided the best tradeoff for the QCAE-SA model (results shown in  section~\ref{sec:results}).
For the inference results and comparisons, the B512 DPU configuration was chosen as the accelerator.

\subsection{Dual-IDS Architecture and Integration}
The LSTM- and autoencoder-based IDSs offer complementary strengths for vehicular network security. 
The LSTM-based IDS, trained via supervised learning, excels at rapidly classifying known attacks such as flooding, fuzzing, and spoofing. 
In contrast, the CAE-based IDS models normal traffic through unsupervised learning, enabling the detection of previously unseen deviations from normal CAN-ID sequence patterns. 
Their integration yields a comprehensive detection architecture: the LSTM-IDS provides precise recognition of established patterns, while the CAE-IDS enhances adaptability to novel behaviours, all with minimal overhead when jointly deployed on a unified platform.

\begin{figure}[t!]
  \centering
  \includegraphics[scale = 1]{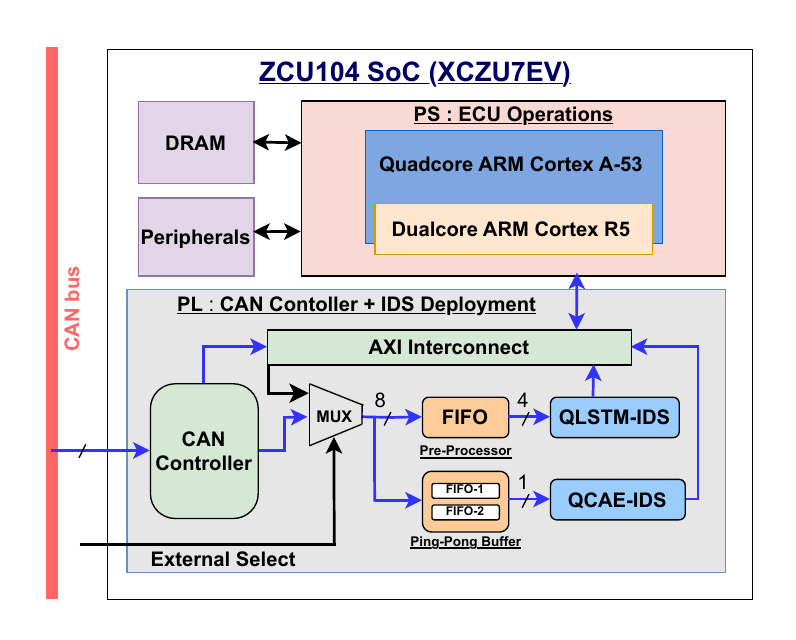}
  \caption{The figure illustrates the ECU architecture modelled on a Zynq Ultrascale+ hybrid SoC and the path of a CAN message upon reception in this architecture. In this architecture, the processing system (PS) functions as the ECU, executing automotive applications. The blue arrows indicate the path taken by the CAN message from the CAN bus to the software tasks on the PS. The IDS  processing is integrated within the programmable logic (PL) section of the XCZU7EV device. The IDSs can receive the CAN message directly from the controller (at the same instance as it is read by the PS) or forwarded by the PS. We use a multiplexer to control this at run-time, with the select line connected to an external peripheral on the board. }
  \label{fig:architecture}
\end{figure}

Figure~\ref{fig:architecture} illustrates the proposed architecture that integrates the dual-IDS function closer to the network interface to reduce software overheads and latency while retaining software-based control.
The Zynq Ultrascale+ XCZU7EV device is used as the target ECU platform, which combines a programmable logic (PL) fabric (a standard FPGA fabric) with an integrated ARM-based CPU subsystem on the same die. 
The ARM subsystem (called processing system - PS) in the XCZU7EV uses a combination of a quad-core ARM Cortex-A53 processor for general processing and a dual-core ARM Cortex-R5 processor for handling real-time tasks.
In our model architecture, the ARM cores on the PS function as the primary ECU compute nodes, responsible for executing standard functionalities, while the programmable logic (PL) serves as an extended compute region. 
The PL enables the deployment of hardware-accelerated IDS models (QLSTM + QCAE) and the CAN network interface, with custom interface designs and low-level optimisations (in case of the models) to exploit hardware-level parallelism and efficiency.
This integration approach also provides a clear separation of tasks between ECU applications and IDS operations. 
By isolating the IDS in the PL, critical ECU compute resources remain dedicated to their primary tasks, ensuring smooth and efficient ECU performance without interference from the IDS.
An open-source CAN controller is used to implement the CAN protocol in this design, with adapted interface wrappers to support the AXI interface to the ARM cores. 

For the IDS integration, we tap into the AXI interconnect between the CAN controller and the PS to drive decoded CAN message data into the IDS engines and the PS at the same time. 
This way, the message data is cloned into the pre-processing FIFO for the QLSTM model and the ping-pong buffer for the QCAE model, when the PS reads the received message from the CAN controller. 
This design is suitable even for proprietary CAN controllers where internal modifications are not feasible.
The pre-processing FIFO of the QLSTM model receives byte-wide message data and converts it to the \textbf{uint4} format required by the model using an asymmetric-width interface. 
In contrast, the pre-processing FIFO of the QCAE model receives the CAN ID of each message in two consecutive bytes, which are stored in a dual-bank ping-pong FIFO buffer.
The pre-processing logic reformats the byte-wide data into the bitwise input format required by the QCAE model.
%\sout{Each byte is passed to a shift register that expands each 8-bit value into its individual bits which are then fed to the model as it's input.}}
%\todo[inline]{If it is reading from the AXI bus, you cannot get bit-by-bit data - what do you mean here? Is it that the QCAE requires each bit as a feature, while the LSTM uses a byte-wide format?\textcolor{blue}{I have corrected this above.}}
This buffer design enables concurrent data acquisition and IDS processing, ensuring no messages are lost during runtime.
Streaming interfaces between the FIFOs and the IDS models. %\todo[inline]{This again contradicts the above - if QCAE is bit-by-bit, then you cannot use a stream of 1 bit? \textcolor{blue}{I think now the above text is correct.}}
The FIFOs also facilitate clock domain crossing from the slower clock domain of the CAN controller to the faster clock domain of the QLSTM and QCAE accelerators.
For hardware-in-the-loop testing, we augment the design with a software-controlled datapath that allows software routines to inject messages directly into the IDS via the AXI interconnect. 
A multiplexer switches the data source between the CAN controller and the AXI interconnect, controlled by an external signal mapped to a physical switch on the ZCU104 board.
Once either IDS model reaches a decision, the result is communicated back to the PS over AXI. 
Based on this output, the ECU in the PS can take appropriate countermeasures, such as dropping malicious messages or triggering safety protocols, thereby enabling real-time defence against known and unknown attacks.
The two IDS components operate concurrently and their decisions are combined through an OR-based alert policy. 
QLSTM-IDS generates a classification for each CAN frame, whereas QCAE-IDS generates an anomaly decision for each block of 100 CAN identifiers. 
An alert is raised when either detector produces a positive decision. 
A QLSTM-only alert is assigned the attack class predicted by the supervised model, while a QCAE-only alert is reported as an unclassified anomaly. 
When both detectors produce positive decisions, the QLSTM class is retained and the QCAE output provides corroborating anomaly evidence. Neither detector overrides or suppresses the other: a benign QLSTM decision does not cancel a QCAE anomaly, and a benign QCAE decision does not cancel a QLSTM attack classification. 
This policy allows the payload-aware QLSTM and ID-sequence-based QCAE to provide complementary coverage.

\section{Results}\label{sec:results}
In this section, we present the performance metrics of the QLSTM-IDS and QCAE-IDS models on the evaluated attack datasets, including accuracy, inference latency, energy consumption, and resource utilisation, on the Zynq Ultrascale+ XCZU7EV platform. 
% Additionally, we provide a discussion on the use of QNNs as IDSs, as explored in existing research literature. 
% \todo[inline]{Is this the case? Would it not be better in the background or in a discussion section?\textcolor{blue}{I commented the above line as it is already part of the background section.}}
%As mentioned, we use the Zynq Ultrascale+ XCZU7EV SoC device (on the ZCU104 development board) which features real-time and high-performance ARM cores in the PS and a programmable logic on the same die as our emulated ECU.
To demonstrate the classification performance of the QLSTM-IDS model, we test it with both the open-car hacking \& survival analysis datasets.
We compare the model performance in terms of accuracy metrics precision, recall and F1 score against the following state-of-the-art IDSs proposed in the research literature: BIDS~\cite{rangsikunpum2024bids}, QMLP-IDS~\cite{khandelwal2023exploring}, DCNN~\cite{song2020vehicle}, TCAN-IDS~\cite{cheng2022tcan}, GRU~\cite{ma2022gru}, HybridSecNet~\cite{10556797}, LSTM~\cite{hossain2020lstm}, TAN-IDS~\cite{10141599}, NLP~\cite{balasubramanian2024canlp}, Random Forest~\cite{santablack}.
To demonstrate the classification performance of the QCAE-IDS model we test it with the survival analysis dataset.
We compare the model performance in terms of accuracy metrics, precision, recall and F1 score against the following state-of-the-art IDSs proposed in the research literature: LSTM~\cite{hossain2020lstm}, TAN-IDS~\cite{10141599}, NLP~\cite{balasubramanian2024canlp}, Random Forest~\cite{santablack}.
We also quantify the detection latency, resource utilisation \& actual energy consumption of our QLSTM and QCAE models on the FPGA and compare the results with other works which have reported these results. 
We used a combination of FINN and Vitis (HLS + Synthesis) flow to generate the streamlined accelerator for the QLSTM model, whereas the QCAE model was mapped to the B512 DPU using the Vitis-AI flow. 
The IDS components are integrated with the CAN controller and the PS through the IP-integrator flow in Vivado and subsequently synthesised to generate the bitstream for the dual-IDS solution. 
The QLSTM accelerator uses a clock frequency of 150\,MHz, whereas the DPU IP for the QCAE is clocked at 300\,MHz.
%, which is deployed on the logic fabric (programmable logic) section of the Zynq XCZU7EV FPGA with a clock frequency of 150\,MHz. 
%We used Vitis-AI and Vivado to generate the bitstream for the QCAE-IDS, which is deployed on the logic fabric (programmable logic) section of the Zynq XCZU7EV FPGA with a clock frequency of 300\,MHz. 
We use the PYNQ image on the ARM cores, which uses a custom Ubuntu OS with PetaLinux tools since it offers an easy interface to manage data movement to and from the accelerators on the PL for extensively testing the inference performance of our models. 
The heavy OS layers allow us to compute the worst-case latency in the presence of software and driver overheads.
Additionally, APIs in the PYNQ libraries allow us to isolate power measurement of the logic region by directly monitoring the power rails from the ARM cores while the models run inference.

\subsection{Accuracy}
\subsubsection{QLSTM-IDS - Message Level Detection}
\begin{figure}[htbp]
    \centering

    % Top row: one figure
    \begin{subfigure}{0.5\textwidth}
        \centering
        \includegraphics[width=0.7\textwidth]{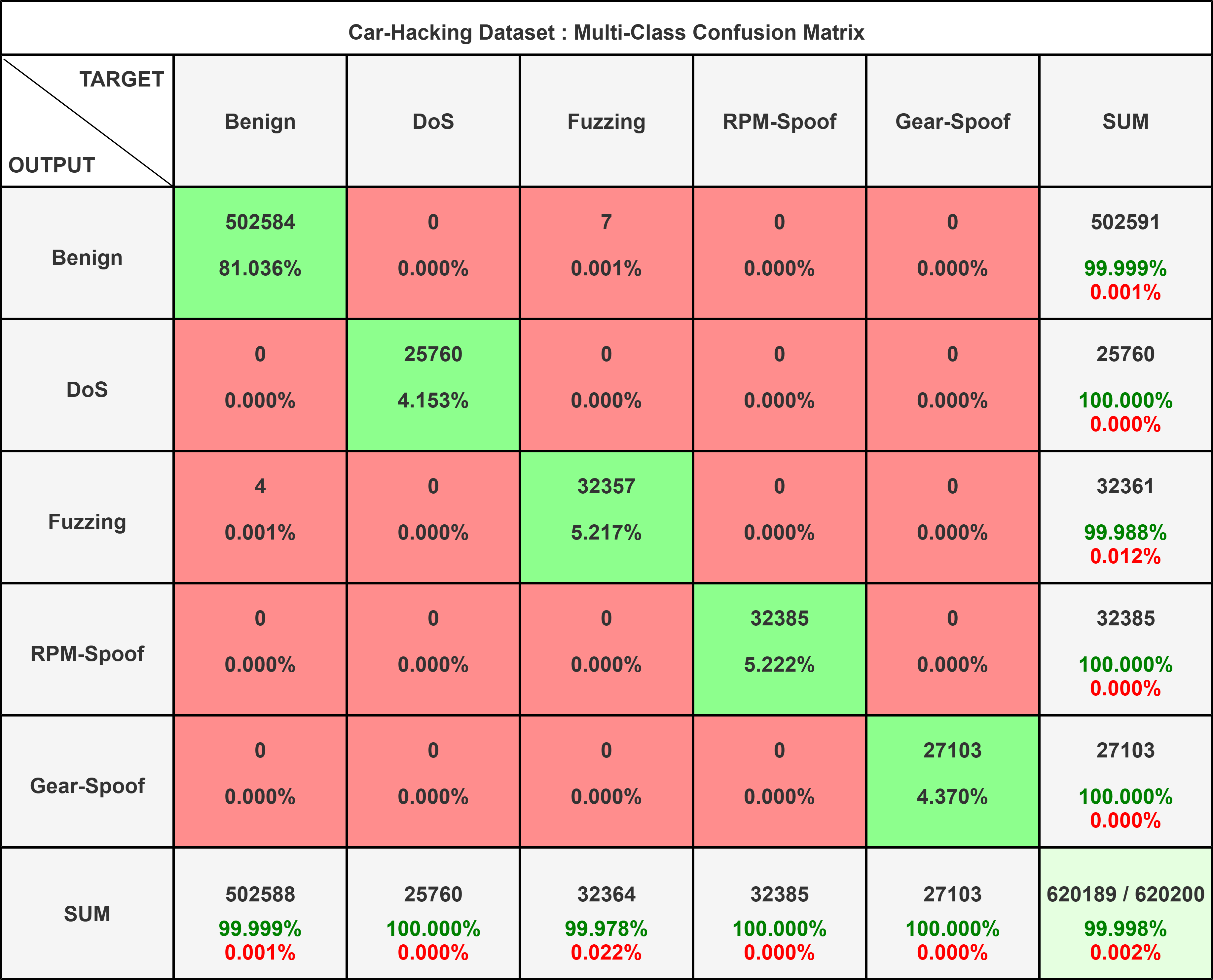}
        \caption{Car Hacking Dataset confusion matrix}
        \label{fig:subfig2}
    \end{subfigure}

    \vspace{0.5em} % space between rows

    % Bottom row: three side-by-side figures
    \begin{subfigure}{0.32\textwidth}
        \centering
        \includegraphics[width=\textwidth]{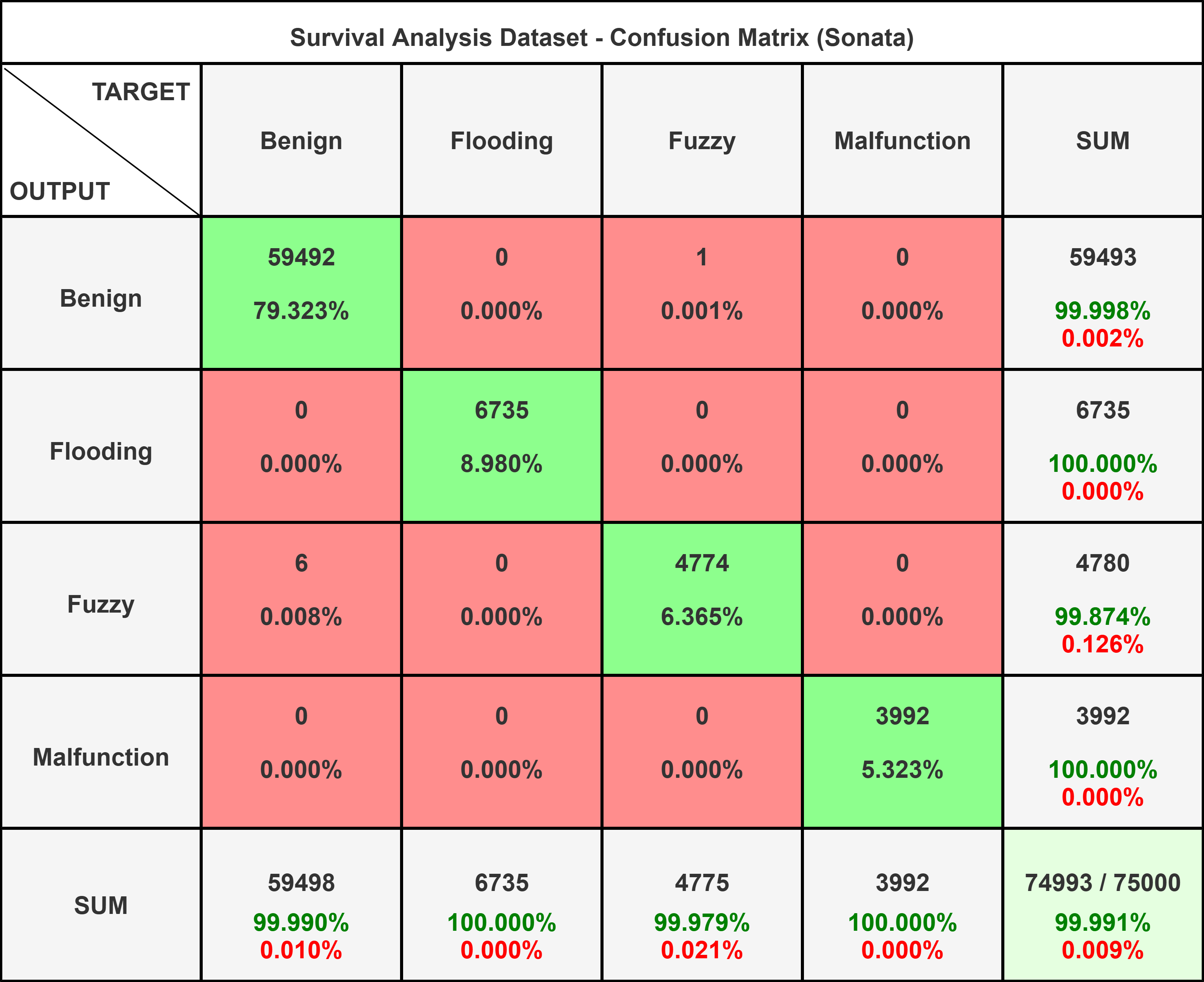}
        \caption{HYUNDAI YF Sonata 2010 Model}
        \label{fig:subfig1}
    \end{subfigure}
    \hfill
    \begin{subfigure}{0.32\textwidth}
        \centering
        \includegraphics[width=\textwidth]{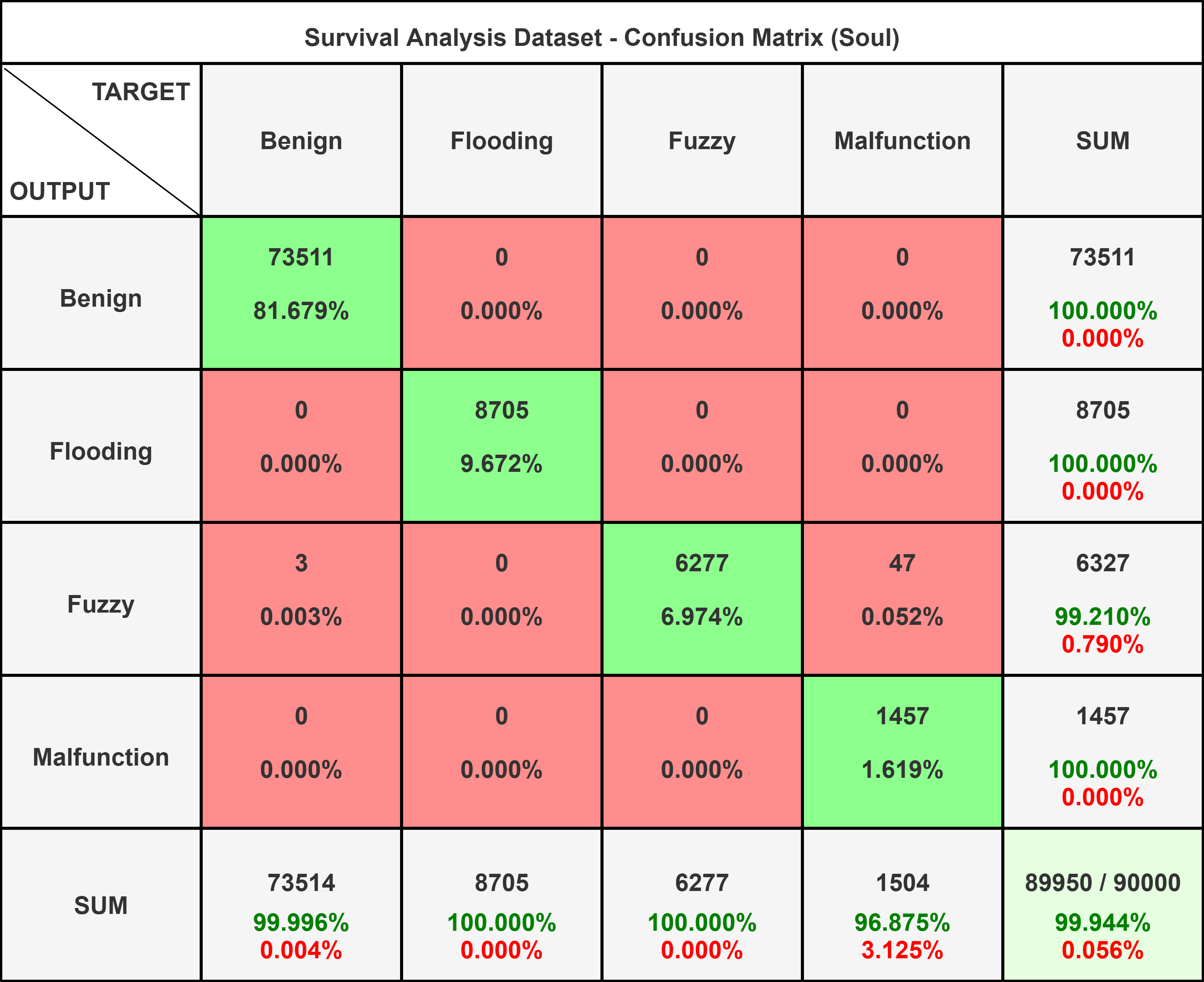}
        \caption{KIA compact SUV Soul 2015 Model}
        \label{fig:subfig2}
    \end{subfigure}
    \hfill
    \begin{subfigure}{0.32\textwidth}
        \centering
        \includegraphics[width=\textwidth]{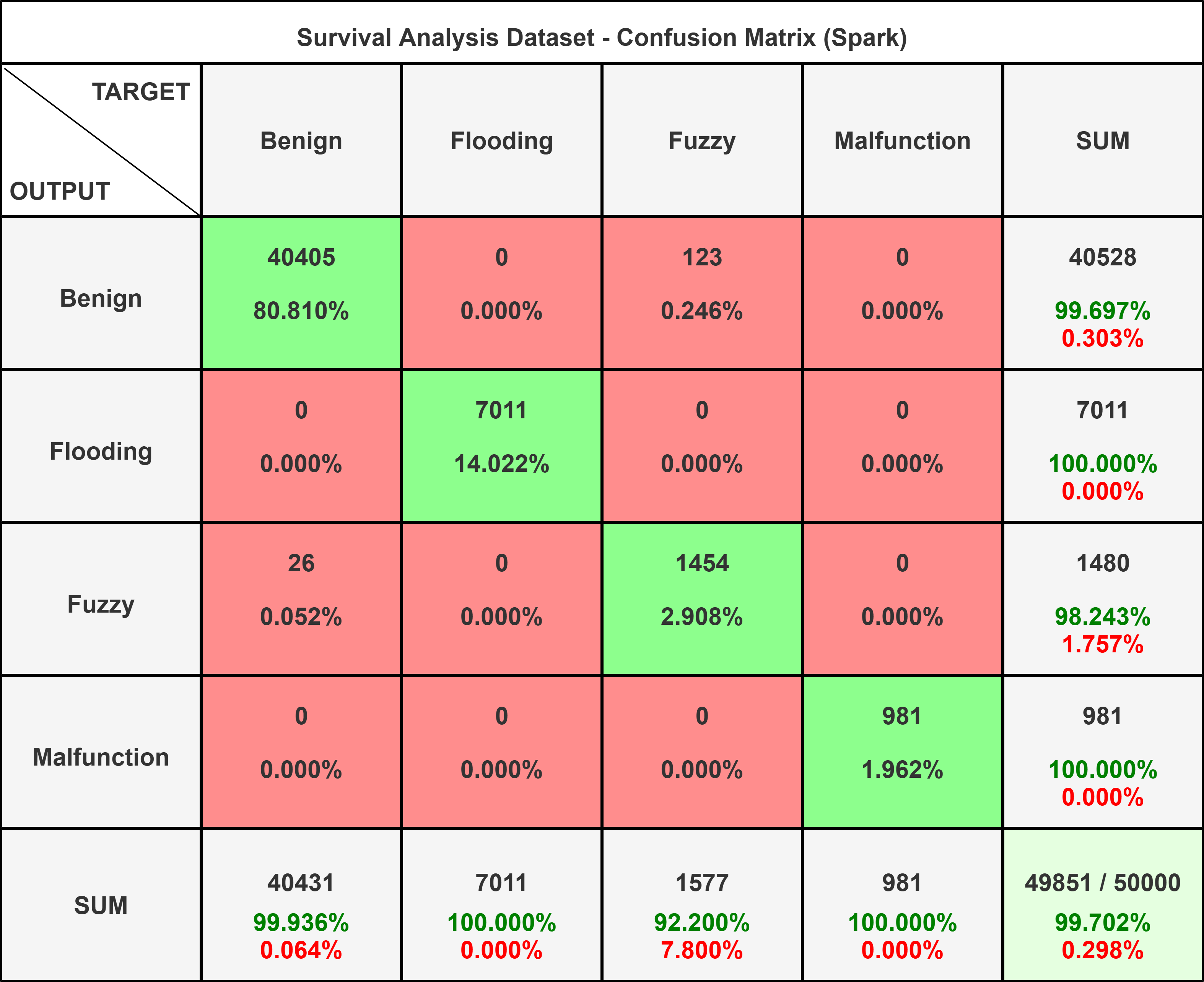}
        \caption{CHEVROLET mini-compact vehicle Spark 2015 Model}
        \label{fig:subfig3}
    \end{subfigure}

    \caption{Confusion matrices for QLSTM-IDS classification results on the car hacking and survival analysis datasets. (Sonata, Soul, and Spark models)}
    \label{fig:cm-ch-sad}
\end{figure}

\textbf{Car Hacking dataset}: On the Car Hacking dataset, our QLSTM-IDS model achieves an impressive overall classification accuracy of 99.998\%. 
The test set includes 620,000 messages, consisting of 502,591 benign, 25,760 DoS, 32,361 Fuzzy, 32,385 RPM-Spoof, and 27,103 Gear-Spoof messages. 
The classified outputs are captured using a multi-class confusion matrix shown in Figure~\ref{fig:cm-ch-sad}, which reveals that the model only misclassified \emph{11} messages across the entire test set.
From the results, we also calculate key performance metrics, including precision, recall, F1 score, and false negative rate (FNR), to capture our model’s detection performance in more detail.
We compare these results with several state-of-the-art IDS models that have used the same dataset for training and testing, including DCNN~\cite{song2020vehicle}, NovelADS~\cite{agrawal2022novelads}, TCAN-IDS~\cite{cheng2022tcan}, GRU~\cite{ma2022gru}, HybridSecNet~\cite{10556797}, BIDS~\cite{rangsikunpum2024bids}, and CQMLP-IDS~\cite{khandelwal2023exploring}.
The comparative results from the referenced studies are presented in Table~\ref{table:sidebyside} with some observations discussed as follows:
\begin{itemize}
    \item \textbf{DoS attack} : QLSTM-IDS model achieves perfect detection, with an F1 score superior to DCNN~\cite{song2020vehicle}, NovelADS~\cite{agrawal2022novelads}, GRU~\cite{ma2022gru}, BIDS~\cite{rangsikunpum2024bids}, and CQMLP-IDS~\cite{khandelwal2023exploring} by margins of 0.05\%, 0.06\%, 0.08\%, 0.37\%, and 0.1\%.
    \item \textbf{Fuzzing attack} : QLSTM-IDS model outperforms DCNN~\cite{song2020vehicle}, TCAN-IDS~\cite{cheng2022tcan}, GRU~\cite{ma2022gru}, BIDS~\cite{rangsikunpum2024bids}, and CQMLP-IDS~\cite{khandelwal2023exploring} in F1 score by 0.2\%, 0.8\%, 0.8\%, 0.3\%, and 0.2\%, respectively, and performs equivalently to NovelADS~\cite{agrawal2022novelads}.
    \item \textbf{RPM-Spoofing} : QLSTM-IDS model achieves perfect detection with F1 score improvement over NovelADS~\cite{agrawal2022novelads}, TCAN-IDS~\cite{cheng2022tcan}, GRU~\cite{ma2022gru}, and BIDS~\cite{rangsikunpum2024bids} by 0.1\%, 0.1\%, 0.4\%, and 0.14\%, respectively.
    \item \textbf{Gear-Spoofing} : QLSTM-IDS model achieves perfect detection with a better F1 score than DCNN~\cite{song2020vehicle}, NovelADS~\cite{agrawal2022novelads}, GRU~\cite{ma2022gru}, and BIDS~\cite{rangsikunpum2024bids} by 0.06\%, 0.1\%, 0.4\%, and 0.1\%.
\end{itemize} 
In comparison to the multi-attack detection model HybridSecNet~\cite{10556797}, which reports average precision, recall, and F1 scores of 99.57\%, 99.58\%, and 99.57\% respectively, the proposed QLSTM-IDS achieves 99.99\% across all three metrics. 
This represents a consistent improvement of +0.4\% in precision, recall, and F1, highlighting it's advantage in multi-attack detection scenarios. 

\begin{table}[t!]
\centering
\caption{Accuracy metric comparison (\%) of our QLSTM-IDS accelerator against the reported literature on the DoS, Fuzzing \& Spoofing (RPM \& Gear) attacks for car hacking dataset.}

\begin{minipage}{0.49\textwidth}
\centering
\scalebox{0.75}{
\begin{tabular}{@{}llrrrr@{}}
\toprule
\multirow{2}{*}{\textbf{Attack}}  & \multirow{2}{*}{\textbf{Model}} & {\textbf{Precision}} & {\textbf{Recall}} & {\textbf{F1}}  & \textbf{FNR} \\
 & & \textbf{(\%)} & \textbf{(\%)} & \textbf{(\%)} & \textbf{(\%)}\\
\midrule
\multirow{5}{*}{DoS} 
& DCNN~\cite{song2020vehicle}                  & 100                & 99.89          & 99.95  & 0.13     \\
& NovelADS~\cite{agrawal2022novelads}        & 99.97               & 99.91          & 99.94  & -     \\
& TCAN-IDS~\cite{cheng2022tcan}             & 100             & 99.97          & 99.98  & -     \\
& GRU~\cite{ma2022gru}                  & 99.93             & 99.91               & 99.92  & -     \\
& BIDS~\cite{rangsikunpum2024bids}     & 99.50   &   97.73  & 98.63  & -     \\
& QMLP-IDS~\cite{khandelwal2023exploring} &  99.92            &   99.88       &   99.90  &  0.11     \\
& \textbf{QLSTM-IDS}                  &    100          &       100         & 100  &   0   \\
\midrule
\multirow{5}{*}{Fuzzy} 
& DCNN~\cite{song2020vehicle}                 & 99.95             & 99.65          & 99.80  & 0.5     \\
& NovelADS~\cite{agrawal2022novelads}                  & 99.99               & 100         & 100  & -     \\
& TCAN-IDS~\cite{cheng2022tcan}                  & 99.96             & 99.89          & 99.22  & -     \\
& GRU~\cite{ma2022gru}                  & 99.32             & 99.13               & 99.22  & -     \\
& BIDS~\cite{rangsikunpum2024bids}     &  100  &  99.38   & 99.69  & -     \\
& QMLP-IDS~\cite{khandelwal2023exploring} &   99.93     & 99.69          & 99.81 &  0.27      \\
& \textbf{QLSTM-IDS}            &  99.99            &     99.98      &  99.98  &  0.02    \\
\bottomrule
\end{tabular}}
\end{minipage}
\hfill
\begin{minipage}{0.49\textwidth}
\centering
\scalebox{0.75}{
\begin{tabular}{@{}llrrrr@{}}
\toprule
\multirow{2}{*}{\textbf{Attack}}  & \multirow{2}{*}{\textbf{Model}} & {\textbf{Precision}} & {\textbf{Recall}} & {\textbf{F1}}  & \textbf{FNR} \\
 & & \textbf{(\%)} & \textbf{(\%)} & \textbf{(\%)} & \textbf{(\%)}\\
 \midrule
\multirow{5}{*}{RPM} 
& DCNN~\cite{song2020vehicle}                  & 99.99 &   99.94 & 99.96  &  0.11    \\
& NovelADS~\cite{agrawal2022novelads}          &    99.91 & 99.9  & 99.91  &      \\
& TCAN-IDS~\cite{cheng2022tcan}                  &   99.9 &  99.9  & 99.9  &      \\
& GRU~\cite{ma2022gru}                  &     99.95   &  99.31   & 99.63  &      \\
& BIDS~\cite{rangsikunpum2024bids}     & 99.99   &  99.73   & 99.86  & -     \\
& QMLP-IDS~\cite{khandelwal2023exploring} &   99.96 &   100       & 99.98     &  0     \\
& \textbf{QLSTM-IDS}                  & 100             & 100               & 100  & 0     \\
\midrule
\multirow{5}{*}{Gear} 
& DCNN~\cite{song2020vehicle}                 &     99.99 &    99.89       &  99.94 & 0.12     \\
& NovelADS~\cite{agrawal2022novelads}       &    99.89  &    99.93  &  99.91 & -     \\
& TCAN-IDS~\cite{cheng2022tcan}           &     99.99         &  99.98         & 99.98  & -     \\
& GRU~\cite{ma2022gru}                  &     99.95 &    99.31  & 99.63  & -     \\
& BIDS~\cite{rangsikunpum2024bids}     & 99.99   &  99.84    &  99.91  & -     \\
& QMLP-IDS~\cite{khandelwal2023exploring} &      -       &       -    &  - & -     \\
& \textbf{QLSTM-IDS}            & 100             & 100          & 100   &  0    \\
\bottomrule
\end{tabular}}
\end{minipage}

\label{table:sidebyside}
\end{table}

% \begin{table}[t!]
%     \centering
%     \caption{Confusion matrix capturing the classification results of the QLSTM-IDS for the Car Hacking Dataset.}
%     \scalebox{1}{% Adjust scale if needed
%         \begin{tabular}{llcrrrr}
%             \toprule
%             \multicolumn{1}{c}{} & \multicolumn{5}{c}{\textbf{Predicted Classes}} \\
%             \cmidrule{2-6}
%             \textbf{Actual Class} & \textbf{Normal} & \textbf{DoS} & \textbf{Fuzzing} & \textbf{Spoofing-RPM} & \textbf{Spoofing-Gear} \\ 
%             \midrule
%             \textbf{Normal}         & 502584 &  0  &  7   &  0   &   0   \\ 
%             \textbf{DoS}            & 0     & 25760 & 0     & 0    & 0     \\ 
%             \textbf{Fuzzing}        & 4     & 0    & 32357 & 0    & 0     \\ 
%             \textbf{Spoofing-RPM}   & 0      & 0    & 0     & 32385 & 0     \\ 
%             \textbf{Spoofing-Gear}  & 0      & 0    & 0     & 0    & 27103 \\ 
%             \bottomrule
%         \end{tabular}
%     }
%     \label{table:cm-qlstm-car-hacking}
% \end{table}

% \begin{figure}[t!]
%   \centering
%   \includegraphics[scale = 0.1]{figures/CM-CHD.png}
%   \caption{Confusion matrix capturing the classification results of the QLSTM-IDS for the Car Hacking Dataset. The figure also shows the success/failure rates for individual attacks as well as overall performance numbers.}
%   \label{fig:cm-chd}
% \end{figure}

\textbf{Survival Analysis Dataset :} The survival analysis dataset, as discussed in the previous section, includes attack datasets for three different car models: HYUNDAI Sonata (2010), KIA Soul SUV (2015), and  CHEVROLET mini/compact vehicle Spark (2015). 
We evaluate QLSTM-IDS against the attack datasets from each of these vehicle models. 
The test set for each model consists of the following number of messages: Sonata - 75,000 (Normal: 59,493, Flooding: 6,735, Fuzzing: 4,780, Malfunction: 3,992); Soul - 90,000 (Normal: 73,511, Flooding: 8,705, Fuzzing: 6,327, Malfunction: 1,457); and Spark - 50,000 (Normal: 40,528, Flooding: 7,011, Fuzzing: 1,480, Malfunction: 981). 

% \begin{figure}[htbp]
%     \centering
%     \begin{subfigure}{0.46\textwidth}
%         \centering
%         \includegraphics[width=\textwidth]{figures/CM-CHD.png}
%         \caption{Car Hacking Dataset confusion matrix}
%         \label{fig:subfig2}
%     \end{subfigure}
%     \begin{subfigure}{0.48\textwidth}
%         \centering
%         \includegraphics[width=\textwidth]{figures/CM-Sonata.png}
%         \caption{HYUNDAI YF Sonata 2010 Model}
%         \label{fig:subfig1}
%     \end{subfigure}
%     % \hfill
%     \begin{subfigure}{0.48\textwidth}
%         \centering
%         \includegraphics[width=\textwidth]{figures/CM-Soul.png}
%         \caption{KIA compact SUV Soul 2015 Model}
%         \label{fig:subfig2}
%     \end{subfigure}
%     \begin{subfigure}{0.48\textwidth}
%         \centering
%         \includegraphics[width=\textwidth]{figures/CM-Spark.png}
%         \caption{CHEVROLET mini-compact vehicle Spark 2015 Model}
%         \label{fig:subfig3}
%     \end{subfigure}
%     \caption{Confusion matrices for QLSTM-IDS classification results on the car hacking and survival analysis datasets. (Sonata, Soul, and Spark models)}
%     \label{fig:cm-ch-sad}
% \end{figure}

The multi-class confusion matrices generated from the classification output of our model across different vehicle models in the survival analysis dataset are shown in Figure~\ref{fig:cm-ch-sad}.
From the results, we can observe that our model output produces 7, 50, and 149 mis-classifications across the test sets for Sonata, Soul, and Spark cars, respectively. 
We compare our findings with state-of-the-art models that have utilized this dataset, including XGBoost~\cite{anjum2022vehicle}, LSTM~\cite{hossain2020lstm}, G-IDCS~\cite{park2023g}, TAN-IDS~\cite{10141599}, NLP~\cite{balasubramanian2024canlp}, and Random Forest~\cite{santablack}.
Table~\ref{table:combined_sa_performance} shows the comparative performance of our QLSTM-IDS against the above state-of-the-art models across the three vehicle datasets, Sonata, Soul and Spark respectively and we discuss some observations from the comparisons below: 
\begin{itemize}
\item \textbf{Sonata} : QLSTM-IDS achieves a 5.3\% higher F1 score than XGBoost~\cite{anjum2022vehicle} while being on par with the other two models in detecting \emph{flooding} attacks. 
In the case of detecting fuzzing and malfunction attacks, QLSTM-IDS improves on XGBoost~\cite{anjum2022vehicle} and Random Forest~\cite{santablack} by 0.4\% \& 0.07\% (\emph{fuzzing}), and by 0.04\% \& 0.06\% (\emph{malfunction}) respectively, while being nearly identical to the LSTM~\cite{hossain2020lstm} model. 
\item \textbf{Soul} :  QLSTM-IDS for the \emph{flooding} attack achieves on par performance with Random Forest~\cite{santablack}, LSTM~\cite{hossain2020lstm} and Transformer-based IDS~\cite{10141599}, while improving on the XGBoost model by 5.3\% (F1 score).
For the \emph{fuzzing} attack, the model improved on the LSTM~\cite{hossain2020lstm} and Transformer~\cite{10141599} models by 2.6\% and 2.3\% (F1 scores) respectively, while slightly falling behind XGBoost and Random Forest by 0.2\% and 0.3\% (F1 scores).
For \emph{malfunction attacks}, the model outperforms all the other models with perfect detection with zero misclassification, improving on XGBoost, Random Forest, LSTM and Transformer models by 3.5\%, 0.06\%, 9.1\%, and 0.2\% (F1 scores), respectively.
\item \textbf{Spark}: For \emph{flooding} and \emph{malfunction} attacks, the model achieves perfect detection with zero misclassification across the test data.
However,for the \emph{fuzzing} attack, the model achieves good precision but with degraded recall, F1 scores and FNRs, producing better results than only the XGBoost model. 
We believe that this degradation stems from the very low number of test data available in the Spark subset and specifically for the \emph{fuzzing} attack on this vehicle.  
\end{itemize}
Additionally, in~\cite{balasubramanian2024canlp}, the authors report an average F1 score of 97.6\% for the IDS model proposed in the paper, when averaged across all attacks and all vehicles. 
In comparison, QLSTM-IDS achieves average F1 scores of 99.97\%, 99.86\%, and 98.37\% across the attack datasets for the \emph{Sonata}, \emph{Soul}, and \emph{Spark} models, respectively, with an overall average of 99.4\%.
It should be noted that all competing methods (except XGBoost) use a block of acquired CAN data as the input feature for detecting the attack, with adapted labels that flag the block as malicious; in contrast, our model performs detection on a per-message basis and flags only the specific message as an attack or benign. 

\begin{table*}[t!]
\centering
\caption{Performance metric comparison of QLSTM-IDS across different attacks for Sonata (left), Soul (middle), and Spark (right) datasets.}
\label{table:all-metrics-3col}

\begin{tabular}{@{}ccc@{}}
% Row 1
\begin{minipage}{0.31\textwidth}
\centering
\scriptsize
\begin{tabular}{lccl}
\toprule
\textbf{Model} & Prec. & Recall & F1 \\
\midrule
XGBoost~\cite{anjum2022vehicle} & 100 & 90 & 94.74 \\
RF~\cite{santablack} & 100 & 100 & 100\\
G-IDCS~\cite{park2023g} & 99.73 & 99.72 & 99.72 \\
LSTM~\cite{hossain2020lstm} & - & 100 & 100 \\
\textbf{QLSTM-IDS} & 100 & 100 & 100 \\
\bottomrule
\end{tabular}
\caption*{(a) Sonata - Flooding}
\end{minipage}
&
\begin{minipage}{0.31\textwidth}
\centering
\scriptsize
\begin{tabular}{lccl}
\toprule
\textbf{Model} & Prec. & Recall & F1 \\
\midrule
XGBoost~\cite{anjum2022vehicle} & 100 & 90 & 94.74 \\
RF~\cite{santablack} & 100 & 100 & 100 \\
LSTM~\cite{hossain2020lstm} & - & 100 & 100 \\
G-IDCS~\cite{park2023g} & 96.92 & 99.74 & 98.31 \\
%Transformer~\cite{10141599} & 100 & 100 & 100  \\
\textbf{QLSTM-IDS} & 100 & 100 & 100 \\
\bottomrule
\end{tabular}
\caption*{(b) Soul - Flooding}
\end{minipage}
&
\begin{minipage}{0.31\textwidth}
\centering
\scriptsize
\begin{tabular}{lccl}
\toprule
\textbf{Model} & Prec. & Recall & F1\\
\midrule
            XGBoost~\cite{anjum2022vehicle} & 100 & 90 & 94.74  \\
            RF~\cite{santablack} & 100 & 100 & 100 \\
            LSTM~\cite{hossain2020lstm} & - & 100 & 100 \\
            G-IDCS~\cite{park2023g} & 99.65 & 100 & 99.83 \\
            %Transformer~\cite{10141599} & 100 & 100 & 100 \\
            \textbf{QLSTM-IDS} & 100 & 100 & 100 \\
\bottomrule
\end{tabular}
\caption*{(c) Spark - Flooding}
\end{minipage}
\\[1.5em]

% Row 2
\begin{minipage}{0.31\textwidth}
\centering
\scriptsize
\begin{tabular}{lccl}
\toprule
\textbf{Model} & Prec. & Recall & F1\\
\midrule
XGBoost~\cite{anjum2022vehicle} & 99.98 & 99.08 & 99.53 \\
RF~\cite{santablack} & 99.79 & 99.93 & 99.86 \\
G-IDCS~\cite{park2023g} & 100 & 100 & 100 \\
LSTM~\cite{hossain2020lstm} & - & 99.95 & 99.96  \\
\textbf{QLSTM-IDS} & 99.87 & 99.98 & 99.93  \\
\bottomrule
\end{tabular}
\caption*{(d) Sonata - Fuzzing}
\end{minipage}
&
\begin{minipage}{0.31\textwidth}
\centering
\scriptsize
\begin{tabular}{lccl}
\toprule
\textbf{Model} & Prec. & Recall & F1\\
\midrule
XGBoost~\cite{anjum2022vehicle} & 99.99 & 99.58 & 99.78\\
RF~\cite{santablack} & 99.79 & 99.93 & 99.86 \\
LSTM~\cite{hossain2020lstm} & - & 94.69 & 97.01\\
G-IDCS~\cite{park2023g} & 100 & 99.84 & 99.92 \\
%Transformer~\cite{10141599} & 96.84 & 97.84 & 97.33 \\
\textbf{QLSTM-IDS} & 99.21 & 100 & 99.60 \\
\bottomrule
\end{tabular}
\caption*{(e) Soul - Fuzzing}
\end{minipage}
&
\begin{minipage}{0.31\textwidth}
\centering
\scriptsize
\begin{tabular}{lccl}
\toprule
\textbf{Model} & Prec. & Recall & F1 \\
\midrule
XGBoost~\cite{anjum2022vehicle} & 97.34 & 92 & 94.6\\
RF~\cite{santablack} & 99.79 & 99.93 & 99.86 \\
LSTM~\cite{hossain2020lstm} & - & 98.23 & 98.31 \\
G-IDCS~\cite{park2023g} & 98.92 & 100 & 99.46 \\
%Transformer~\cite{10141599} & 98.85 & 99.39 & 99.12 & - \\
\textbf{QLSTM-IDS} & 98.24 & 92.20 & 95.13 \\
\bottomrule
\end{tabular}
\caption*{(f) Spark - Fuzzing}
\end{minipage}
\\[1.5em]

% Row 3
\begin{minipage}{0.31\textwidth}
\centering
\scriptsize
\begin{tabular}{lccl}
\toprule
\textbf{Model} & Prec. & Recall & F1  \\
\midrule
XGBoost~\cite{anjum2022vehicle} & 99.92 & 100 & 99.96 \\
RF~\cite{santablack} & 99.90 & 99.98 & 99.94  \\
G-IDCS~\cite{park2023g} & 100 & 99.64 & 99.82 \\
LSTM~\cite{hossain2020lstm} & - & 100 & 100  \\
\textbf{QLSTM-IDS} & 100 & 100 & 100 \\
\bottomrule
\end{tabular}
\caption*{(g) Sonata - Malfunction}
\end{minipage}
&
\begin{minipage}{0.31\textwidth}
\centering
\scriptsize
\begin{tabular}{lccl}
\toprule
\textbf{Model} & Prec. & Recall & F1  \\
\midrule
XGBoost~\cite{anjum2022vehicle} & 100 & 93.22 & 96.49  \\
RF~\cite{santablack} & 99.90 & 99.98 & 99.94  \\
LSTM~\cite{hossain2020lstm} & - & 96.97 & 90.87  \\
G-IDCS~\cite{park2023g} & 95.92 & 99.03 & 97.45  \\
%Transformer~\cite{10141599} & 99.65 & 99.95 & 99.80 \\
\textbf{QLSTM-IDS} & 100 & 100 & 100 \\
\bottomrule
\end{tabular}
\caption*{(h) Soul - Malfunction}
\end{minipage}
&
\begin{minipage}{0.31\textwidth}
\centering
\scriptsize
\begin{tabular}{lccl}
\toprule
\textbf{Model} & Prec. & Recall & F1  \\
\midrule
XGBoost~\cite{anjum2022vehicle} & 99.80 & 100 & 99.90\\
RF~\cite{santablack} & 99.90 & 99.98 & 99.94 \\
LSTM~\cite{hossain2020lstm} & - & 100 & 100\\
G-IDCS~\cite{park2023g} & 99.36 & 100 & 99.68 \\
% Transformer~\cite{10141599} & 99.17 & 99.78 & 99.47 & - \\
\textbf{QLSTM-IDS} & 100 & 100 & 100 \\
\bottomrule
\end{tabular}
\caption*{(i) Spark - Malfunction}
\end{minipage}
\end{tabular}
\label{table:combined_sa_performance}
\end{table*}

\subsubsection{QCAE-IDS - Block Level Detection}
To test the functional correctness of the quantised QCAE-SA model, we initially compare the pre- \& post-quantisation inference performance of the model. 
For the SA dataset, we observe an equal performance for the DoS \& Fuzzing attacks with a slight drop for the Malfunction attack. 
The QCAE-SA model performance is shown in table~\ref{table:qcae_perf_sa_combined}. 
The model is not evaluated on the Soul subset of the dataset because it contained insufficient attack-free messages to support meaningful training, validation, and threshold calibration using separate benign subsets.
This limited the development of a reliable UL model for this subset for detecting previously unseen CAN-ID sequence anomalies.

\begin{table}[t!]
\centering
\caption{Accuracy metric comparison (\%) of our QCAE-IDS accelerator pre- and post-quantisation on the Sonata and Spark subsets of the SA dataset.}
\scalebox{0.9}{
\begin{tabular}{@{}lllllll@{}}
\toprule
\textbf{Subset} & \textbf{Attack} & \textbf{Model} & \textbf{Precision} & \textbf{Recall} & \textbf{F1} \\
\midrule
\multirow{6}{*}{Sonata} 
& \multirow{2}{*}{Flooding}     & Pre-Q     & 100     & 99.28   & 99.64 \\
&                                   & QCAE-SA   & 100     & 99.28   & 99.64 \\
& \multirow{2}{*}{Fuzzy}        & Pre-Q     & 100     & 99.5    & 99.75 \\
&                                   & QCAE-SA   & 100     & 99.5    & 99.75 \\
& \multirow{2}{*}{Malfunction}  & Pre-Q     & 100     & 96.8    & 98.37 \\
&                                   & QCAE-SA   & 100     & 94.24   & 97.04 \\
\midrule
\multirow{6}{*}{Spark}  
& \multirow{2}{*}{Flooding}     & Pre-Q     & 98.92   & 100     & 99.33 \\
&                                   & QCAE-SA   & 98.21   & 100     & 98.88 \\
& \multirow{2}{*}{Fuzzy}        & Pre-Q     & 100     & 98.39   & 99.19 \\
&                                   & QCAE-SA   & 100     & 98.39   & 99.19 \\
& \multirow{2}{*}{Malfunction}  & Pre-Q     & 98.75   & 93.89   & 95.89 \\
&                                   & QCAE-SA   & 97.08   & 94.53   & 94.99 \\
\bottomrule
\end{tabular}
}
\label{table:qcae_perf_sa_combined}
\end{table}

The 150 benign blocks per vehicle were used for threshold calibration. 
Their zero observed false positives are a calibration result and are not an independent estimate of test-set false-positive performance.
To evaluate detection of attacks unseen by QCAE during model development, we evaluated QCAE-SA on unseen \emph{Flooding}, \emph{Fuzzing}, and \emph{Malfunction} attacks from the SA dataset.
We also compare QCAE-SA with published IDS methods including supervised classifiers trained on labelled benign and attack samples.
Table~\ref{table:confmatrix_cae_sa} captures the classification performance of the model in isolation across the test set as a confusion matrix. 
As discussed in section~\ref{subsec:cae-ids}, the accuracies reported are obtained for the threshold values of \texttt{73} \& \texttt{52} for the QCAE-SA model on the Sonata and Spark test sets.
The misclassifications (false negatives) observed while testing on unseen attack datasets can be attributed to scenarios where the attack message flow closely resembles that of the normal messages within the observed block of messages at a given time.

% \begin{table}[t!]
% \centering
% \caption{\textcolor{blue}{Confusion matrix capturing the classification results of the QCAE-SA on the attack-free (benign), Flooding, Fuzzing, and Malfunction attacks for the SA dataset. (Note the model is trained only on benign messages and does not see any attack message during the training phase.)}}
%     \scalebox{1}{
%         \begin{tabular}{@{}llrr@{}}
%         \toprule
%             \textbf{Attack} & \textbf{Message Type}  & Predicted Normal  & Predicted Attack  \\ 
%             \midrule
%           \multirow{2}{*}{Benign} & True Normal & 150 & 0\\ 
%             & True Attack & NA & NA \\
%           \multirow{2}{*}{Flooding} & True Normal & 585 & 0\\ 
%             & True Attack & 3 & 412 \\ 
%            %\midrule
%             \multirow{2}{*}{Fuzzy} & True Normal & 606 & 0  \\
%            &  True Attack & 2 & 392 \\ 
           
%            \multirow{2}{*}{Malfunction} & True Normal & 594 & 0 \\
%            &  True Attack & 24 & 393 \\ 
%             \bottomrule
%         \end{tabular}}
% \label{table:confmatrix_cae_sa}
% \end{table}

\begin{table}[t!]
\centering
\caption{Confusion matrix capturing the classification results of the QCAE-SA on the Flooding, Fuzzing, and Malfunction attacks for the SA dataset (Sonata \& Spark). (Note the model is trained only on benign messages and does not see any attack message during the training phase.)}
\begin{minipage}{0.49\textwidth}
\centering
% \caption{\textcolor{blue}{Inference accuracy metrics of CAE-SA pre- and post-quantisation (QCAE-SA) on all the attacks of the SA dataset.}} %\todo{I could 
\scalebox{0.9}{
\begin{tabular}{@{}llrr@{}}
\toprule
\textbf{Attack} & \textbf{Message Type}  & Pred. N  & Pred. A  \\ 
\midrule
          \multirow{2}{*}{Flooding} & True N & 585 & 0\\ 
            & True A & 3 & 412 \\ 
           %\midrule
            \multirow{2}{*}{Fuzzy} & True N & 606 & 0  \\
           &  True A & 2 & 392 \\ 
           
           \multirow{2}{*}{Malfunction} & True N & 594 & 0 \\
           &  True A & 24 & 393 \\ 
\bottomrule
\end{tabular}}
\end{minipage}
\hfill
\begin{minipage}{0.49\textwidth}
\centering
\scalebox{0.9}{
\begin{tabular}{@{}llrr@{}}
\toprule
\textbf{Attack} & \textbf{Message Type}  & Pred. N  & Pred. A  \\ 
 \midrule
          \multirow{2}{*}{Flooding} & True N & 548 & 10\\ 
            & True A & 0 & 442 \\ 
           %\midrule
            \multirow{2}{*}{Fuzzy} & True N & 464 & 3  \\
           &  True A & 0 & 183 \\ 
           
           \multirow{2}{*}{Malfunction} & True N & 465 & 14 \\
           &  True A & 17 & 294 \\ 
\bottomrule
\end{tabular}}
\end{minipage}
\label{table:confmatrix_cae_sa}
\end{table}

% \textcolor{blue}{The inference performance of the QCAE-SA model is compared against the following IDSs proposed in the literature: XGBoost~\cite{anjum2022vehicle}$^{\mathrm{S}}$, Random Forest~\cite{santablack}$^{\mathrm{S}}$, G-IDCS~\cite{park2023g} and LSTM~\cite{hossain2020lstm}$^{\mathrm{S}}$ and discussed below (Note, the three methods marked with $\mathrm{S}$ use supervised learning with labelled benign and attack samples. QCAE-SA is trained exclusively on benign data.}):
The inference performance of the QCAE-SA model is compared against the following IDSs proposed in the literature: XGBoost~\cite{anjum2022vehicle}, Random Forest~\cite{santablack}, LSTM~\cite{hossain2020lstm} and G-IDCS~\cite{park2023g}, with the first three methods using supervised learning with labelled benign and attack samples, as opposed to the QCAE-SA which was trained exclusively on benign data.
\begin{itemize}
    \item \textbf{Sonata}: For the \emph{flooding} attack, the QCAE-SA model performs better than~\cite{anjum2022vehicle} by $\approx$ 5\% in terms of F1  score, while being slightly less than~\cite {santablack,hossain2020lstm} by 0.4\% and almost equal to~\cite{park2023g}.
    For the \emph{fuzzing} attack, the QCAE-SA model performs better than~\cite{anjum2022vehicle} by 0.2\% in terms of F1 score, while being slightly less than~\cite{santablack,hossain2020lstm,park2023g} by 0.1\%, 0.2\% \& 0.2\% respectively. 
    For the \emph{malfunction} attack, the QCAE-SA model suffers from a higher FNR, which leads to it being less than~\cite{anjum2022vehicle,santablack,hossain2020lstm,park2023g} by 3\% in terms of F1 score.
    \item \textbf{Spark}: For the \emph{flooding} attack, the QCAE-SA model performs better than~\cite{anjum2022vehicle} by $\approx$ 4\% in terms of F1  score, while being slightly less than ~\cite{santablack,hossain2020lstm,park2023g}by 1\%.
    For \emph{fuzzing} attacks, the QCAE-SA model performs better than~\cite{anjum2022vehicle} by 4.5\% while being slightly less than~\cite{santablack,hossain2020lstm,park2023g, 10141599} by 0.7\%, 0.1\%, 0.2\% \& 0.1\% respectively.
    For \emph{malfunction} attacks, the QCAE-SA model suffers from a higher FNR, which leads to it being less than~\cite{anjum2022vehicle,santablack,hossain2020lstm,park2023g} by 5\% in terms of F1 score.
\end{itemize}
The loss in performance for the malfunction attack can be attributed to the fact that these messages resemble the normal ECU communication closely.
The detailed results of this model against these works can be found in Table~\ref {table:comp_sa} \& Table~\ref{table:comp_sa_spark} respectively.

While evaluated independently due to differing input granularity and dataset availability, QLSTM-IDS and QCAE-IDS are complementary.
Together, they enable a dual IDS architecture: QLSTM-IDS ensures accurate detection of known attacks at the message level, while QCAE-IDS adds robustness against novel or unseen attacks at the block level.

\begin{table*}[t!]
\centering
\begin{minipage}{0.45\textwidth}
    \centering
    \caption{Accuracy metric comparison (\%) of the QCAE-SA model for the Sonata subset of SA dataset against the reported literature on the Flooding, Fuzzing and Malfunction attacks.}
    \scalebox{0.75}{
    \begin{tabular}{@{}lllll@{}}
    \toprule
    \textbf{Attack}  & \textbf{Model} & \textbf{Precision} & \textbf{Recall} & \textbf{F1} \\
    \midrule
    \multirow{4}{*}{\textbf{Flooding}} & XGBoost~\cite{anjum2022vehicle} & 100  &  90  & 94.74   \\ 
    & RF~\cite{santablack} & 100 & 100 &100 \\
    & G-IDCS~\cite{park2023g} & 99.72 & 99.72 & 99.72 \\
    & LSTM~\cite{hossain2020lstm} & - & 100 & 100 \\
    & \textbf{QCAE-SA} & \textbf{100} & \textbf{99.28} & \textbf{99.64} \\
    \midrule
    \multirow{4}{*}{\textbf{Fuzzing}} & XGBoost~\cite{anjum2022vehicle} & 99.98 & 99.08 & 99.53 \\ 
    & RF~\cite{santablack} & 99.79 & 99.93 & 99.86\\
    & G-IDCS~\cite{park2023g} & 100 & 100 & 100 \\
    & LSTM~\cite{hossain2020lstm} & - & 99.95 & 99.96 \\
    & \textbf{QCAE-SA} & \textbf{100} & \textbf{99.5} & \textbf{99.75} \\
    \midrule
    \multirow{4}{*}{\textbf{Malfunction}} & XGBoost~\cite{anjum2022vehicle} & 99.92 & 100 & 99.96  \\ 
    & RF~\cite{santablack} & 99.90 & 99.98 & 99.94 \\
    & G-IDCS~\cite{park2023g} & 100 & 99.64 & 99.82 \\
    & LSTM~\cite{hossain2020lstm} & - & 100 & 100 \\
    & \textbf{QCAE-SA} & \textbf{100} & \textbf{94.24} & \textbf{97.04} \\
    \bottomrule
    \end{tabular}}
    \label{table:comp_sa}
\end{minipage}
\hfill
\begin{minipage}{0.45\textwidth}
    \centering
    \caption{Accuracy metric comparison (\%) of the QCAE-SA model for the Spark subset of SA dataset against the reported literature on the Flooding, Fuzzing and Malfunction attacks.}
    \scalebox{0.75}{
    \begin{tabular}{@{}lllll@{}}
    \toprule
    \textbf{Attack}  & \textbf{Model} & \textbf{Precision} & \textbf{Recall} & \textbf{F1} \\
    \midrule
    \multirow{5}{*}{\textbf{Flooding}} & XGBoost~\cite{anjum2022vehicle} & 100 & 90 & 94.74\\ 
    & RF~\cite{santablack} & 100 & 100 & 100 \\
    & LSTM~\cite{hossain2020lstm} & - & 100 & 100 \\
    & G-IDCS~\cite{park2023g} & 99.65 & 100 & 99.83 \\
    % & TAN-IDS & 100 & 100 & 100  \\
    & \textbf{QCAE-SA} & \textbf{98.21} & \textbf{100} & \textbf{98.88}  \\
    \midrule
    \multirow{5}{*}{\textbf{Fuzzing}} & XGBoost~\cite{anjum2022vehicle} & 97.34 & 92 & 94.6  \\ 
    & RF~\cite{santablack} & 99.79 & 99.93 & 99.86  \\
    & LSTM~\cite{hossain2020lstm} & - & 98.23 & 98.31  \\
    & G-IDCS~\cite{park2023g} & 98.92 & 100 & 99.46 \\
    % & TAN-IDS & 98.85 & 99.39 & 99.12  \\
    & \textbf{QCAE-SA} & \textbf{100} & \textbf{98.39} & \textbf{99.19}  \\
    \midrule
    \multirow{5}{*}{\textbf{Malfunction}} & XGBoost~\cite{anjum2022vehicle} & 99.80 & 100 & 99.90 \\ 
    & RF~\cite{santablack} & 99.90 & 99.98 & 99.94 \\
    & LSTM~\cite{hossain2020lstm} & - & 100 & 100 \\
    & G-IDCS~\cite{park2023g} & 99.36 & 100 & 99.68 \\
    % & TAN-IDS & 99.17 & 99.78 & 99.47 \\
    & \textbf{QCAE-SA} & \textbf{97.08} & \textbf{94.53} & \textbf{94.99}  \\
    \bottomrule
    \end{tabular}}
    \label{table:comp_sa_spark}
\end{minipage}
\end{table*}

\subsection{Detection Latency}

One of the principal challenges in deploying IDSs in in-vehicle networks is the processing latency associated with complex detection models. 
To examine the real-time viability of our proposed low-precision QLSTM-IDS and QCAE-IDS accelerators, we integrated both directly into the CAN network controller and conducted detailed hardware based latency measurements. 
In each case, the controller’s internal buffers were pre-loaded with representative CAN messages from the evaluation datasets, and latency was measured from the point of data availability at the controller interface to the generation of classification results. 
The reported latency comprises (i) IDS inference time per message for QLSTM-IDS and per block for QCAE-IDS and (ii) the Zynq processor’s readout time for transferring results from the accelerators. 
Timing commences when the controller asserts the data-valid signal. 
Two measurement flows were considered:
\begin{itemize}
    \item \textbf{Direct interface path}: Data is delivered directly from the CAN controller interface to the IDS hardware chain, thereby isolating hardware latency.
    \item \textbf{AXI-DMA path}: CAN messages are transferred via AXI-DMA from the PYNQ interface, such that latency also includes data movement overheads. In this case, the timer is triggered immediately prior to asserting the IDS enable input, and the multiplexer is configured to the DMA input.
\end{itemize}
Both flows inherently capture operating system and driver overheads introduced by the PYNQ image on the ARM cores. 
Minimal variation was observed across the two setups, and all results were averaged across multiple runs to mitigate OS- and software-level jitter. 
The QLSTM-IDS accelerator achieved an end-to-end latency of $\approx$ 0.25 ms per message, corresponding to a throughput of $\approx$ 4,000 messages/s. 
The block-based QCAE-IDS achieved a latency of $\approx$ 0.42 ms per block of 100 frames. 
For the latter, a ping-pong buffering scheme in programmable logic ensures that one buffer accumulates incoming frames while the other is processed, effectively concealing inference latency within block reception and enabling real-time operation under high-load CAN traffic.
The two accelerators are implemented as independent, spatially allocated hardware pipelines and therefore do not require time-multiplexed execution or software-level scheduling between the QLSTM-IDS and QCAE-IDS. 
Each incoming CAN message is forwarded to the QLSTM-IDS processing path while its CAN ID is concurrently accumulated in the active QCAE-IDS ping-pong buffer.

When benchmarked against prior work, these results underscore the efficiency of the proposed approach as shown in Table~\ref{table:latcomp}. 
The QLSTM-IDS delivers a 2.29$\times$ reduction in latency compared with MTH-IDS~\cite{yang2021mth}, the most directly comparable per-message technique. While our earlier work on quantised MLPs~\cite{khandelwal2022light,khandelwal2023exploring} achieved lower raw latency due to reduced computational complexity, those methods relied on sliding window block processing, whereas QLSTM-IDS provides true per-message granularity, a critical requirement for real-time intrusion detection. 
Likewise, the quantised QCAE-SA yields a 27.6$\times$ latency improvement relative to the unsupervised Transformer IDS of~\cite{10141599}, while offering performance nearly identical to that of an int8 CAE deployed on the Jetson Xavier GPU (0.41~ms). 
The reported results were obtained at an accelerator frequency of 150~MHz; however, the relatively high overheads of the PYNQ read/write APIs indicate that further reductions are achievable under bare-metal or RTOS deployments. 
Collectively, these findings demonstrate that QLSTM-IDS achieves fine-grained, low-latency intrusion detection, whereas QCAE-IDS provides efficient block-level detection with strong robustness to unseen attacks. 
Both approaches substantially outperform or match contemporary software- and hardware-based IDS implementations, confirming their suitability for latency-critical CAN environments.
Since automotive systems employ heterogeneous computing platforms with substantially different capabilities, the reported latency values should be interpreted in the context of their respective target applications, input granularities, and measurement protocols. 
Accordingly, the comparison illustrates the deployment-level implications of different IDS integration approaches rather than providing a direct platform-normalised performance benchmark.

% \todo[inline]{I think there could be better coherence here - it is as if we are benchmarking two different solutions. Maybe we can combine the measurements of our models in one paragraph and then move the comparisons to the second paragraph? \textcolor{blue}{This section is now reorganised.}}

\begin{table}[t]
\centering
\caption{Detection latency comparison of the proposed IDS against other state-of-the-art IDSs reported in literature.}
\scalebox{1}{%0.9
\begin{tabular}{@{}lrll@{}}
\toprule
\textbf{Models}     & \textbf{Latency} & \textbf{Frames} & \textbf{Platform} \\ \cmidrule{1-4} 
GRU~\cite{ma2022gru} & 890\,ms & 5000 CAN frames & Jetson Xavier NX  \\%GPU-128\\
MLIDS~\cite{desta2020mlids} & 275\,ms & per CAN frame & GTX Titan X \\%GPU-3072\\
%Rec-CNN~\cite{desta2022rec} & 117\,ms & 128 CAN frames & Jetson TX2 \\%GPU-128\\
NovelADS~\cite{agrawal2022novelads} & 128.7\,ms & 100 CAN frames & Jetson Nano \\%GPU-128\\
%GIDS~\cite{seo2018gids} & 5.89\,ms & 64 CAN frames &  GTX 1080  \\%GPU-2560 \\ 
TAN-IDS~\cite{10141599} & 11.6\,ms & 128 CAN frames & - \\
DCNN~\cite{song2020vehicle} &  5\,ms & 29 CAN frames & Tesla K80 \\%GPU-2496      \\
TCAN-IDS~\cite{cheng2022tcan} & 3.4\,ms & 64 CAN frames & Jetson AGX \\%GPU-512 \\
MTH-IDS~\cite{yang2021mth} &  0.574\,ms & per CAN frame & Raspberry Pi 3     \\
QMLP-IDS~\cite{9974508} &  0.24\,ms & per CAN frames & Zynq Ultrascale+     \\
BIDS~\cite{rangsikunpum2024bids} &  0.17\,ms & 48 CAN frames & Zynq     \\
CQMLP-IDS~\cite{khandelwal2023exploring} & 0.11 ms & per CAN frames & Zynq Ultrascale+    \\
\textbf{QCAE-IDS} & 0.42 ms & 100 CAN frames & Zynq Ultrascale+    \\ 
\textbf{QLSTM-IDS} & 0.25 ms & per CAN frame & Zynq Ultrascale+    \\ 
\bottomrule
\end{tabular}}
\label{table:latcomp}
\end{table}

% \begin{table}[ht]
% \centering
% \caption{Detection latency comparison based on processing type: per CAN frame (QLSTM-IDS) vs. message block (QCAE-IDS).}
% \begin{minipage}{0.48\textwidth}
% \centering
% \scalebox{0.95}{
% \begin{tabular}{@{}lrl@{}}
% \toprule
% \textbf{Model} & \textbf{Platform} & \textbf{Latency} \\
% \midrule
% MLIDS~\cite{desta2020mlids} & GTX Titan X & 275\,ms  \\
% MTH-IDS~\cite{yang2021mth} & Raspberry Pi 3  & 0.574\,ms \\
% QMLP-IDS~\cite{9974508} & Zynq Ultrascale+& 0.24\,ms   \\
% CQMLP-IDS~\cite{khandelwal2023exploring} & Zynq Ultrascale+ & 0.11\,ms  \\
% \textbf{QLSTM-IDS}& Zynq Ultrascale+ & 0.25\,ms  \\
% \bottomrule
% \end{tabular}}
% \caption*{(a) Per CAN Frame Processing}
% \end{minipage}
% \hfill
% \begin{minipage}{0.49\textwidth}
% \centering
% \scalebox{0.95}{
% \begin{tabular}{@{}lrll@{}}
% \toprule
% \textbf{Model} & \textbf{Platform} & \textbf{Frames} & \textbf{Latency} \\
% \midrule
% GRU~\cite{ma2022gru}  & Jetson Xavier NX   & 5000& 890\,ms\\
% NovelADS~\cite{agrawal2022novelads} & Jetson Nano  & 100& 128.7\,ms \\
% TAN-IDS~\cite{10141599} & -   & 128 & 11.6\,ms \\
% DCNN~\cite{song2020vehicle}  & Tesla K80  & 29& 5\,ms \\
% TCAN-IDS~\cite{cheng2022tcan} & Jetson AGX   & 64 & 3.4\,ms \\
% BIDS~\cite{rangsikunpum2024bids} & Zynq  & 48  & 0.17\,ms \\
% \textbf{QCAE-IDS} & Zynq Ultrascale+   & 100 & 0.42\,ms\\
% \bottomrule
% \end{tabular}}
% \caption*{(b) Block-Based Processing}
% \end{minipage}
% \label{tab:latency_tables}
% \end{table}

\subsection{Energy \& Resource Utilisation}
To evaluate the energy consumption of the dual-IDS system, we employ the \emph{pynq-rails} feature from the PYNQ-PMBus package, which can be used to monitor and isolate the power rails of the processor subsystem (ARM-cores) and programmable logic (logic region) on the Zynq Ultrascale+ device.
To measure the energy cost of the IDSs, we monitor and isolate any power rails that power the logic region of the Zynq, while triggering the IDS to perform inference on a test dataset. 
The measurements are averaged over 10000 messages and inference operations, with each new message arriving at the peak rate (4000 messages per second) determined from the latency measurement.
At this operating point, the energy consumption is 0.8~mJ per QLSTM-IDS frame-level inference and 1.1~mJ per QCAE-IDS block-level inference, corresponding to 1.9~mJ when each accelerator performs one inference. These values are of the same order of magnitude as the reported energy consumption of GRU~\cite{ma2022gru} (1.77~mJ) and MTH-IDS~\cite{yang2021mth} (1.29~mJ) per inference respectively.
While this consumption is marginally higher than the energy consumed by other quantised MLP models, our dual-IDS model represents a better trade-off including:
\begin{itemize}
    \item multi-class attack classification with higher accuracy
    \item benign-trained detection of CAN-ID sequence anomalies, and
    \item improved generalisability across diverse attack modes and datasets.
\end{itemize}
We further assess the resource utilisation of the  dual-IDS accelerator and the CAN network controller (CAN-NC), with metrics gathered post-implementation via Vitis-HLS/Vivado. 
As detailed in Table~\ref{table:resource-utilisation}, the accelerator uses less than 40\% of LUTs and under 15\% of flip-flops (FFs), leaving substantial resources available for additional accelerators or applications on the FPGA. 
This efficient resource utilisation highlights the design’s scalability and flexibility, allowing other concurrent workloads to be offloaded from the ECU to hardware blocks on the logic region, without affecting IDS functionality.
Table~\ref{table:dpucomparison} presents a comparison between the optimised QCAE model and the prior implementation from~\cite{khandelwal2023real}, referred to as QCAE-CH, in terms of resource utilisation, inference latency, and energy consumption.
The proposed QCAE-IDS achieves significant improvements on the XCZU7EV device, reducing LUT usage by 10\%, FFs by 14\%, DSPs by 33\%, BRAM by 19\% and URAM usage by 35\% respectively.
Additionally, it achieves a 1.57$\times$ reduction in active power consumption while maintaining the same processing latency as the QCAE-CH model.
Since the QLSTM-IDS and QCAE-IDS models occupy separate, statically allocated FPGA resources and use independent I/O paths, they operate concurrently without sharing compute resources or introducing runtime resource contention.
\begin{table}[t]
\caption{The resource consumption overheads of QLSTM-IDS, the QCAE-IDS and the CAN Controller on the PL region of the Zynq XCZU7EV SoC.} 
\begin{center}
\scalebox{1}{%
\begin{tabular}{@{}lccccc@{}}
\toprule
\textbf{Function} & \textbf{LUTs} (\%) & \textbf{FFs} (\%) & \textbf{LUTRAM} (\%) & \textbf{BRAMs} (\%) & \textbf{DSPs} (\%)\\
\midrule
QLSTM-IDS & 40902 (17.75) & 11234 (2.44) & 4669 (4.59) & 23.5 (7.53) & 12 (0.69) \\
QCAE-IDS & 39956 (17.34) & 51957 (11.1) & - & 50 (16.51) & 125 (7.18) \\
CAN-NC & 887 (0.38) & 625 (0.14) & 18 (0.02) & -& - \\
\midrule
\textbf{Overall} & 81745 (35.47) & 63816 (13.68) & 4687 (4.61) & 73.5 (24.04)& 137 (7.87) \\
\bottomrule
\end{tabular}}
\label{table:resource-utilisation}
\end{center} 
\end{table}

% \textcolor{blue}{The power consumption and resource overhead of the approach is measured to quantify the energy consumption per inference and the hardware resources incurred on the device in table~\ref{table:dpucomparison}.
% The table shows the deployment of the models on three separate DPU configurations. 
% It is observed that the QCAE-SA model consumes \& 2.74\,W when measured directly from the device's power rails (using the PYNQ-PMBus package) while performing inference and other tasks on the ECU (with Linux OS), which translates to 1.1\,mJ of energy consumed per inference by the models.
% Among other reported results in the literature, the QCAE-SA model results in a $\approx$ 2$\times$ reduction in power consumption when compared to the GRU~\cite{ma2022gru} model, which uses an Nvidia Jetson Xavier as a dedicated IDS node.
% It is also observed the per-inference energy consumption of the INT8 QCAE-SA models on the Jetson Xavier to be 1.7\.mJ when averaged over 10000 inference runs.
% The integrated ECU approach with QCAE-IDS achieves a 1.5$\times$ reduction in energy consumed per inference for the QCAE-SA model. 
% Hence, the model is either identical or better in terms of latency and energy utilisation over a dedicated Jetson Xavier-based IDS accelerator, making the proposed approach a more appropriate choice for deploying IDS in critical CAN systems. 

\begin{table}[t!]
\centering
\caption{Performance comparison of the QCAE-CH \& the proposed QCAE-IDS model based on per message latency (ms), resource (\%), \& power consumption (W).}
\scalebox{1}{
\begin{tabular}{@{}lcccccccc@{}}
\toprule
\multirow{2}{*}{\textbf{Model}} & \multicolumn{1}{c}{\textbf{Latency (ms)}} & \multicolumn{5}{c}{\textbf{\% Resource Utilisation}} & \multicolumn{2}{c}{\textbf{Power (W)}} \\ \cmidrule{2-9}
 &   & LUT & FF & DSP & BRAM & URAM & Idle & Active\\ \midrule
% B1152           & 0.77 &    0.33           &  19.53&13.92&13.08&19.55&33.33   & 3.12   &  3.56 &  3.33      \\
\textbf{QCAE-CH (B4096)}           & 0.43           &  27.17&25.03&40.74&35.42&47.92   & 4.36   & 4.89    \\
\textbf{QCAE-IDS (B512)}           &   0.42             &  17.34 & 11.10 &7.18 &16.51 &12.5  & 2.61 & 3.11  \\
\bottomrule
\end{tabular}}
\label{table:dpucomparison}
\end{table}

\subsection{Discussion : QNNs proposed as IDSs}
As discussed in the preceding sections, several quantised IDS models have recently been proposed in the research literature to reduce the model complexity and energy consumption.
This subsection examines their accuracy metrics and deployment statistics in comparison to the proposed quantised LSTM-based IDS (QLSTM-IDS), which are shown in Table~\ref{table:qnn-comparison}.
In~\cite{rangsikunpum2024bids}, the authors present a two-stage IDS consisting of a Binarized Neural Network (BNN) and a Generative Adversarial Network (GAN). 
Their model uses a sequence of CAN IDs with a window length of 48 messages as input features. 
However, using only CAN IDs as input features limits the model’s ability to detect attacks that inject malicious payloads without affecting the IDs. 
Additionally, the 1-bit quantised IDS has a higher misclassification rate compared to the proposed QLSTM-IDS.
In~\cite{9974508}, one of our previous works, we propose two separate 8-bit quantised MLP models as IDSs to detect DoS, Fuzzing, and Spoofing (RPM and Gear) attacks. 
The key idea was to optimise the learned features for similar attack vectors, thus improving the individual detector's performance while reducing the model complexity. %\todo[inline]{Is this the right idea? \textcolor{blue}{Yes, sounds reasonable!}}
We observe that the proposed QLSTM-IDS achieves a lower misclassification rate compared to the tailored models while reducing the energy consumption per inference by over 2 $\times$.
Similarly, in ~\cite{khandelwal2023exploring}, we explored extreme quantisation using a 2-bit custom quantised MLP (CQMLP-IDS) to perform multi-class classification to detect DoS, Fuzzing, and RPM-Spoofing attacks. 
This was achieved using a conservative labelling scheme to improve model convergence and to achieve high accuracy, using a sliding window approach over input (and previous) CAN messages. 
With the final implementation, the CQMLP only manages to achieve a slightly higher misclassification rate than the proposed QLSTM-IDS, while the window-based labels and thresholding scheme can lead to dropped benign messages at runtime. 
In~\cite{9895417}, authors propose a fully binarised neural network as an IDS, with very low detection latency of 0.04\,ms.  However, the authors validate the models' performance on a proprietary custom dataset, with the results showing lower accuracy metrics ($\approx$ 90.5\%) for random ID, replay, spoofing and other hybrid attacks in this dataset. 
Compared to other implementations in Table~\ref{table:qnn-comparison}, QLSTM-IDS achieves the lowest misclassification rate (11/620,200), outperforming QMLP, BIDS, and CQMLP by a significant margin. 
It uses moderate hardware resources $\approx$ 17.75\% LUTs and 2.44\% FFs respectively lower than QMLP and BIDS while maintaining minimal BRAM usage (0.82 Mb). 
Although its energy consumption (0.8 mJ) is higher than CQMLP and BIDS, it remains more efficient than QMLP. 
Each inference takes only 0.25 ms, making it well-suited for real-time detection.
Similarly, we previously introduced a unseen attack detection autoencoder, evaluated on the CAR Hacking dataset~\cite{khandelwal2023real}. 
In this work, we proposed QCAE-IDS, an optimised variant of that autoencoder employing filter pruning techniques.
QCAE-IDS is tested on the Survival Analysis dataset, demonstrating robust unknown attack detection capabilities with high accuracy, while also achieving a 1.9$\times$ improvement in per-inference energy efficiency over the previous implementation.
The model also achieves a 27.6$\times$ reduction in processing latency against the other unsupervised learning based transformer based IDS proposed in the literature~\cite{10141599}.
Overall, QLSTM-IDS along with the QCAE-IDS offers the best balance between accuracy, resource utilisation, and deployability among the evaluated models.
Note, the reported classification metrics in the paper are point estimates for a single selected trained variant for both the models and the corresponding fixed data splits.
Repeated seed and alternative split experiments remain part of the future work.

\begin{table}[t]
\caption{This table compares the proposed QLSTM-IDS and QCAE-IDS with other QNN models from the research literature, evaluating them in terms of resource utilization, energy consumption, and overall misclassification rates as reported on the car hacking dataset. The accuracy number for the QCAE-IDS are reported on the Sonata subset of the survival analysis dataset.}
\begin{center}
\scalebox{1}{%
\begin{tabular}{@{}lccccc@{}}
\toprule
\textbf{Model} & \textbf{LUTs} & \textbf{FFs} & \textbf{BRAM} (Mb) & \textbf{Energy} (mJ) & \textbf{Misclassifications} \\
\midrule
QMLP~\cite{9974508}  & 56,733 (24.6\%)& 72,147 (15.6\%) & 3.06 & 1.96 & 75/200,000 \\
CQMLP~\cite{khandelwal2023exploring} & 3999 (1.74\%) & 4524 (0.98\%) & 0.84 & 0.23 & 147/180,000 \\
BIDS~\cite{rangsikunpum2024bids}  & 23,256 (10.09\%) & 37,943 (8.2\%) & 4.07 & 0.35 & 70/187,062  \\
\textbf{QLSTM-IDS} & 40,902 (17.75\%) & 11,234 (2.44\%) & 0.82 & 0.8 & \textbf{11/620,200}\\
\textbf{QCAE-IDS} & 39,956 (17.34\%) & 51,957 (11.1\%) & 1.74 & 1.1 & 29/3011\\
\bottomrule
\end{tabular}}
\label{table:qnn-comparison}
\end{center} 
\end{table}
\section{Conclusion \& Future Work}\label{sec:conclusion} 
In this paper, we proposed a dual-IDS framework tailored for modern automotive networks, addressing both known and unknown attack vectors through supervised and unsupervised learning approaches. 
The first component, a quantised LSTM-Dense model (QLSTM-IDS), is designed to detect multiple attack types including DoS/Flooding, Fuzzing, and Spoofing/Malfunction attacks with high accuracy.
We designed a custom export flow for converting the QLSTM model into the intermediate representation and further developed custom transformations that mapped the compute nodes in the intermediate representations to existing hardware library blocks in the FINN compiler flow, to generate the dataflow accelerator of our model.
The model is trained using quantisation-aware techniques via the Brevitas library, and efficiently mapped to a hardware accelerator using AMD’s FINN toolchain. 
In addition to the supervised QLSTM-IDS, we also introduce an 8-bit quantised convolutional autoencoder (QCAE-IDS) as an unsupervised IDS capable of detecting unknown attacks.
The demonstrated QCAE capability is limited to previously unseen anomalies that alter CAN-ID sequence patterns.
Detection of attacks that modify only payload content while preserving normal identifier sequences is not established in this work.
The hardware accelerator for the QCAE-IDS is generated using the Vitis-AI framework from AMD.
We present a complete deployment pipeline targeting the Zynq Ultrascale+ XCZU7EV (ZCU104) platform, where the IDSs are implemented on the same platform and integrated close to the CAN network interface, minimising software overhead and enabling real-time operation.
QLSTM-IDS is evaluated on two real world public attack datasets namely CAR Hacking and Survival Analysis while QCAE-IDS (an optimised version of our previous implementation) is tested using the survival analysis dataset.
Both the models demonstrate strong performance in terms of detection accuracy, inference latency, and energy efficiency.
Our evaluation confirms that the QLSTM-IDS achieves exceptional average classification accuracies of 99.998\% and 99.90\% across the two datasets, with an energy consumption of only 0.8~mJ per inference and a processing throughput of up to 4,000 CAN messages per second. 
The QCAE model also achieves $>$ 99\% accuracy when tested on the survival analysis dataset (Sonata and Spark vehicles) showing it's efficacy in detecting unknown attacks.
Further, the model achieved a block processing latency of 0.42 ms and energy consumption of only 1.1 mJ per inference. 
The results obtained highlight the suitability of the deployment models for real-time, resource-constrained embedded environments. 
Together the supervised and unsupervised IDS models form a robust, flexible, and deployable solution for securing legacy automotive networks. 
To encourage further research, we have open-sourced our training pipelines, hardware design files, and deployment flows, providing a foundation for future work on efficient IDS architectures for next-generation vehicular systems.
\section{Acknowledgments}
We would like to express our sincere gratitude to Jakoba Petri-König and Thomas Preußer from the AMD research team in Dublin, Ireland, for their invaluable assistance in deploying the quantised version of LSTMs on FPGAs. 
This research was also supported by grants from NVIDIA (academic hardware grant program) and utilised NVIDIA RTX A6000 GPU for model training purposes.
\bibliographystyle{ACM-Reference-Format}
\bibliography{references}

\end{document}